\documentclass[traditabstract]{aa}
\usepackage{psfrag,graphicx}
\usepackage{txfonts}
\usepackage{natbib}
\usepackage{ulem}
\usepackage{color}

\begin{document}
\bibliographystyle{aa}

\title{The impact of magnetic fields on the IMF in star-forming clouds near a supermassive black hole}
\author{S. Hocuk \inst{1}
       \and
       D. R. G. Schleicher \inst{2}
       \and
       M. Spaans \inst{1}
       \and
       S. Cazaux \inst{1}
       }
\institute{Kapteyn Astronomical Institute, University of Groningen,
          P. O. Box 800, 9700 AV Groningen, Netherlands \\
          \email{seyit@astro.rug.nl, spaans@astro.rug.nl, cazaux@astro.rug.nl}
          \and
          Instit\"ut f\"ur Astrophysik, Georg-August-Universit\"at, 
          Friedrich-Hund-Platz 1, 37077 G\"ottingen, Germany \\
          \email{dschleic@astro.physik.uni-goettingen.de}
         }
\titlerunning{The IMF in magnetized clouds near a supermassive black hole}
\authorrunning{Hocuk et al.}
\date{Received \today}

\abstract
{Star formation in the centers of galaxies is thought to yield massive stars with a possibly top-heavy stellar mass distribution. It is likely that magnetic fields play a crucial role in the distribution of stellar masses inside star-forming molecular clouds. In this context, we explore the effects of magnetic fields, with a typical field strength of 38 $\rm \mu G$, such as in RCW 38, and a field strength of 135 $\rm \mu G$, similar to NGC 2024 and the infrared dark cloud G28.34+0.06, on the initial mass function (IMF) near ($\leq 10$ pc) a $\rm 10^{7}$ solar mass black hole. Using these conditions, we perform a series of numerical simulations with the hydrodynamical code FLASH to elucidate the impact of magnetic fields on the IMF and the star-formation efficiency (SFE) emerging from an 800 solar mass cloud. We find that the collapse of a gravitationally unstable molecular cloud is slowed down with increasing magnetic field strength and that stars form along the field lines. The total number of stars formed during the simulations increases by a factor of $1.5-2$ with magnetic fields. The main component of the IMF has a lognormal shape, with its peak shifted to sub-solar ($\rm \leq 0.3 \,M_{\odot}$) masses in the presence of magnetic fields, due to a decrease in the accretion rates from the gas reservoir. In addition, we see a top-heavy, nearly flat IMF above $\sim2$ solar masses, from regions that were supported by magnetic pressure until high masses are reached. We also consider the effects of X-ray irradiation if the central black hole is active. X-ray feedback inhibits the formation of sub-solar masses and decreases the SFEs even further. Thus, the second contribution is no longer visible. We conclude that magnetic fields potentially change the SFE and the IMF both in active and inactive galaxies, and need to be taken into account in such calculations. The presence of a flat component of the IMF would be a particularly relevant signature for the importance of magnetic fields, as it is usually not found in simulations.}

\keywords{Stars: formation -- Stars: mass function -- Magnetohydrodynamics (MHD) -- X-rays: ISM -- Methods: numerical}

\maketitle

\section{Introduction}
Stars are observed to form in molecular cloud fragments (clumps) with typical densities of $\rm n = 10^{4}-10^{6} \,cm^{-3}$. The molecular gas in active environments such as galactic centers is usually turbulent, FWHM (full-width at half-maximum) = $\rm \sim 5\,km/s$, and has temperatures ranging from $\rm 10\,K$ up to a few $\rm 1000\,K$, depending on the local conditions. Molecular clouds are observed to have magnetic fields with strengths varying from $\rm B \sim 10-few\times100 \,\mu G$ \citep{1987A&A...181..119C, 1999ApJ...520..706C, 2001ApJ...554..916B, 2011ApJ...735...64W}. In galactic centers, it is possible that magnetic field strengths are even stronger. From their VLA (Very Large Array) observations, \cite{1987ApJ...320..545Y, 1987ApJ...322..721Y} estimated using dynamical arguments that the field strength in our Galactic center could be as high as $\geq 1000 \,\mu$G. A more recent detection of mG strengths is also reported by \cite{Matsumura2012}. \cite{2010Natur.463...65C} show that the minimum magnetic field strength near the Galactic center is on the order of $\sim50\,\mu$G on 400 pc scales, while they infer a minimum field strength of $100\,\mu$G using different constraints. Probing larger scales, recent observations of the inner kpc of galaxies, \cite{2005A&A...444..739B} and \cite{2007A&A...465..157M} infer magnetic field strengths of $\rm B \sim 60 \,\mu$G in the central star-forming regions of NGC 1097 and NGC 1365, which are amongst the strongest fields detected in spiral galaxies. 

Magnetic fields are an important component of the physics governing cloud evolution. Yet, it is not fully understood what the effects of magnetic fields are during the fragmentation epoch of molecular clouds and during star formation. Such early influences might also affect the distribution of stellar masses and, therefore, the IMF. For a complete understanding of the theory of star formation, it is essential to have a complete insight into the origin of the IMF. According to most studies over the past two decades, it seems that the influence of magnetic fields is quite important during star and structure formation \citep{2001ApJ...555..178F, 2008A&A...487..247F, 2009Sci...324.1408G, 2010ApJ...709...27W, 2010ApJ...720L..26L, 2010A&A...522A.115S, 2011MNRAS.417.1054S, 2011MNRAS.415.1751M}. \cite{2012ApJ...745L..30W} note, for example, that the shape of the filaments of the infrared dark cloud G28.34+0.06 is likely affected by magnetic field lines and speculate that the formation of the filamentary system has been governed by the interplay between a strong magnetic field, self-gravity, and turbulence. Similarly, \cite{2009Sci...324.1408G} find that the collapse of the hot molecular core of the high-mass star-forming region G31.41+0.31 is controlled by the magnetic field, resulting in its hourglass shape. \cite{2010A&A...522A.115S} show that during the formation of the first stars, the magnetic field amplification owing to the small scale dynamo action will amplify the initial weak magnetic field to such an extent that it becomes dynamically relevant in star formation. Their models show that magnetic fields can change the fragmentation properties and also affect the gas accretion rates onto the protostars, but only if the resolution is sufficiently high \citep{2011PhRvL.107k4504F}. Theoretically as well as observationally, magnetic fields are thus shown to play a critical role in the evolution and fragmentation of molecular clouds.

Magnetic fields are generally supposed to slow down collapse because of the added (Lorentz) force, tension, and pressure to the system. A common misconception is that magnetic fields should also increase the characteristic mass of stars in a cluster, because they tend to increase the effective Jeans mass. This is not necessarily true. According to the findings of \cite{2010ApJ...720L..26L}, the magnetic fields rather decrease the characteristic stellar mass in a cluster. These authors note that star formation occurs along the field lines where dense gas accumulates, which increases the density and decreases the Jeans mass, and therefore decreases the characteristic mass of the IMF. While together with many other magnetic field studies, direct turbulent fragmentation is found to be suppressed \citep{2004MNRAS.347.1001H, 2008MNRAS.385.1820P, 2011A&A...528A..72H}. One needs to explore further how these results depend on the chosen initial conditions and the physics of the simulations. Such magnetic field effects could also affect the IMF in star-forming regions near massive black holes. Star-forming clouds in these environments are, however, subject to strong gravitational forces and feedback effects such as irradiation by X-rays in active galactic nuclei (AGN), and if inside massive star-forming regions, to increased UV fluxes and cosmic ray rates \citep{2009ApJ...702...63W, 2010A&A...522A..24H, 2010A&A...518L..42V, 2011A&A...525A.119M, 2011MNRAS.414.1705P}. \cite{2011A&A...536A..41H} showed that the impact of radiative feedback on the IMF is significant. The IMF that arises from an 800 M$_{\odot}$ star-forming cloud which is being irradiated by X-rays, cosmic rays, and UV photons near a black hole is a top-heavy one.

In this paper, we perform a similar study, where we do not investigate stellar feedback, but focus on the effects of magnetic fields in the inner 10 pc of galaxies. For this, we simulate model clouds with typical magnetic field strengths of $\rm B=38 \,\mu G$ and $\rm B=135 \,\mu G$, and a model cloud without magnetic fields. By comparing these models, we evaluate the impact of magnetic fields.

This paper is organized as follows: In section \ref{sec:method}, we introduce the numerical code FLASH, construe the implementation of the magnetohydrodynamics (MHD) module, and describe the initial conditions. In section \ref{sec:collapsingcloud}, we give a detailed description on the cloud's shape and morphology and how these are affected by the black hole's gravity. In section \ref{sec:results}, we present our results on the effects of magnetic fields for the evolution of the cloud and to its star formation. We then look at the resulting IMFs and the SFEs from our models and analyze them. We also consider radiative feedback in the case that an active black hole irradiates the model cloud with X-rays. In section \ref{sec:conclusion}, we discuss our results and present our final conclusions.

\section{Numerical model and simulation setup}
\label{sec:method}
The simulations in this work have been performed using the AMR (adaptive mesh refinement) hydrodynamical code FLASH4 \citep{2000ApJS..131..273F, Dubey2009512}. For modelling the magnetic fields, we make use of the recently freely available `Unsplit Staggered Mesh' MHD scheme \citep[USM,][]{Lee:2009:USM:1486277.1486457}. The USM scheme adopts a dimensionally unsplit integration on a staggered grid, for the multidimensional MHD formulation, based on a finite-volume, higher-order Godunov method. This scheme uses the constrained transport method \citep{1988ApJ...332..659E} to enforce divergence-free constraints of magnetic fields. The USM scheme has shown good performance in comparison to other current state-of-the-art MHD solvers as can be seen from the review paper by \cite{2011ApJ...737...13K}. For solving the Euler equations, we choose to use the HLLD Riemann solver \citep{2005AGUFMSM51B1295M}, which, for supersonic MHD turbulence simulations, is among the most robust and accurate solvers available, while still staying less diffusive than other popular solvers like `Roe' and `HLLC'. We also adopt a monotonized central (mc) slope limiter \citep{VanLeer1977263} to further enhance solution accuracy and stability.

The hydrodynamical equations are solved for ideal, compressible MHD conditions. The mass (continuity), momentum, energy, and induction equation written in conservative form, coupled with the Poisson equation for gravity, are defined as

\begin{equation}
\frac {\partial\rho} {\partial \rm t} + \nabla \cdot (\rho \textbf{v}) = 0,
\label{eq5:euler1}
\end{equation}

\begin{equation}
\frac {\partial\rho\textbf{v}} {\partial \rm t} + \nabla\cdot(\rho \textbf{vv} - \textbf{BB}) + \nabla {\rm P} = \rho \textbf{g},
\label{eq5:euler2}
\end{equation}

\begin{equation}
\rm \frac {\partial E} {\partial t} + \nabla\cdot \left[ (E+P)\textbf{v} - (\textbf{B} \cdot \textbf{v})\textbf{B} \right] = \rho \textbf{g} \cdot \textbf{v},
\label{eq5:euler3}
\end{equation}

\begin{equation}
{\rm \frac {\partial \textbf{B}} {\partial t} + \nabla\cdot(\textbf{vB} - \textbf{Bv})} = 0,
\label{eq5:euler4}
\end{equation}

\noindent
where, $\rho$ is the fluid mass density, \textbf{v} is the fluid velocity vector, \textbf{g} is the gravitational acceleration vector, \textbf{B} is the magnetic field, P is the total pressure, and E is the total specific energy. The total specific energy, E, and the total pressure, P, are defined as

\begin{equation}
{\rm E} = \frac{1}{2} \rho v^2 + \rho \rm E_{int} + \frac{B^2}{8\pi},
\label{eq5:euler5}
\end{equation}

\begin{equation}
\rm P = P_{th} + \frac{B^2}{8\pi},
\label{eq5:euler6}
\end{equation}

\noindent
and the Poisson equation is given by

\begin{equation}
\nabla^{2}\phi = 4\pi {\rm G}\rho  ~~~~\Rightarrow~~~~ \textbf{g} = -\nabla\phi.
\label{eq5:euler7}
\end{equation}

\noindent
In these equations, $\rm E_{int}$ is the internal energy per unit mass, and $\phi$ is the gravitational potential.

\subsection{Particles}
In order to follow individual protostars, high-density gas is converted into Lagrangian (sink) particles, which are then treated as an N-body problem. The sink particles experience forces and may themselves contribute to the dynamics of the system. These are coupled to the gas through the mesh via gravity. Sink particles also have their own motion independent of the grid, so additional equations of motion and forces come into play. Particles are advanced using the second order in time accurate, variable time step Leapfrog method. The forces acting on the particles are handled through

\begin{equation} \rm
m_i \frac {d\textbf{v}_i} {dt} = \textbf{F}_{long,i} + \textbf{F}_{short,i},
\label{eq5:partic1}
\end{equation}

\noindent
where $\rm \textbf{F}_{long,i}$ represents the sum of all long-range forces (coupling all particles) acting on the i-th particle, this is handled through the mesh, and $\rm \textbf{F}_{short,i}$ represents the sum of all short-range forces (coupling only neighboring particles) acting on the particle, which is handled directly between the particles. Only the gravitational force is implemented for the long and short range interactions. Regular particles are embedded in FLASH. Sink particles, on the other hand, are an adapted version of regular particles. The difference between them is that sink particles can grow in mass by accreting gas and by merging with other particles, however, they cannot lose mass or fragment. These particles represent compact objects, in our case protostars. 

Sink particles are created when there is locally an irreversible collapse. The mass of a sink particle at creation depends on the threshold density, which is determined by the Jeans mass. This mass also indirectly depends on the maximum grid resolution, because the size of the local region is based on a choice. Our sink particles typically have a minimum mass of 0.2 solar masses. This will possibly decrease with higher resolution. A sink particle starts to accrete matter immediately after its creation. Accretion accounts for the largest portion of the final star mass. This contribution is estimated to be around 85\%. The mass distribution of stars therefore strongly depends on accretion rather than their mass at creation. The sink particle algorithm used in this study is further explained in detail in  \cite{2010A&A...522A..24H}. This algorithm adopts the same methods as described by \cite{2004ApJ...611..399K} and \cite{2010ApJ...713..269F}.

\subsection{Radiation transport and the EOS}
For our primary cloud models (models A, B, and C), we use isothermal conditions. This is done by using an equation of state of the form $\rm P \propto \rho$ \citep{2010A&A...522A..24H, 2011A&A...536A..41H} and fixing the gas temperature at 10 K. For our cloud models near an active black hole, there is also the presence of X-ray radiation. The transport of X-rays and cooling radiation is handled through a radiative transport code \citep[XDR code,][]{2005A&A...436..397M, 2007A&A...461..793M, 2008ApJ...678L...5S}, which is coupled to FLASH at every time step. For computational speed and numerical efficiency, we make a large set of pre-computed tables with the XDR code and read it into FLASH during initialization. The XDR code incorporates all the heating processes (photo-ionization, yielding non-thermal electrons; FUV pumping, followed by collisional de-excitation), cooling processes (atomic fine-structure and semi-forbidden lines), and molecular transitions (CO, H$_{2}$, H$_{2}$O, OH, and CH). Cosmic rays, dust-gas coupling, and secondary photons from X-rays, like internal UV, are considered as well. The exact implementation of radiative transfer is explained in detail in \citep{2011A&A...536A..41H}.

\subsection{Resolution}
The computational domain encompasses a range of 24 pc in each direction, which has outflow boundaries in terms of space and isolated boundaries in terms of gravity. Our highest resolution is defined by a refinement with a maximum number of 4096$^3$ cells, which yields a minimum scale length of $5.7 \times 10^{-3}$ pc, or about $10^{3}$ AU. We refine the regions according to a Jeans length criterion, where the Jeans length is calculated as

\begin{equation}
\lambda_{\rm J} = \left( \frac {\pi \rm c_{ s}^2} {{\rm G} \rho} \right)^{\frac{1}{2}}.
\label{eq5:jeanslen}
\end{equation}

\noindent
Here, $c_{s}$ is the sound speed, which can be formulated as ${\rm c_{s}^{2}} = \gamma {\rm P}/\rho$ or ${\rm c_{s}^{2}} = \gamma \rm T$ for an ideal gas. $\gamma$ is a parameter which depends on the equation of state (EOS). For a polytropic EOS, the pressure scales as $\rm P = K \rho^{\gamma}$. A $\gamma = 1$ in this case represents an isothermal state. Whenever the Jeans length of any cell in a block of 16$^{3}$ cells would fall below 20 times the cell size, refinement is initiated for the whole block. When $\lambda_{\rm J}$ is resolved by over 50 times the cell size, de-refinement is initiated in order to save computational time and memory.

\subsection{Initial conditions}
Similar to the performed studies in \cite{2010A&A...522A..24H} and \cite{2011A&A...536A..41H}, we simulate a gravitationally unstable spherical molecular cloud at d$_{\rm bh}$ = 10 pc from a 10$^{7}$ M$_{\odot}$ black hole and follow its collapse. The initial cloud has a number density of 10$^{5}$ cm$^{-3}$ with a mean molecular mass $\mu = 2.3$ and a radius of 0.33 pc. These conditions create a cloud of 800 solar masses. The rest of the 24$^3$ pc sized simulation box has a number density of 10 cm$^{-3}$ giving a total mass of 8000 M$_{\odot}$. The cloud is in a Keplerian orbit around the black hole with an orbital velocity of $\sqrt{\rm GM/d_{bh}} = 66.7$ km/s. This introduces some gravitational shear into the cloud. The shearing velocity across the cloud is given by

\begin{equation} \rm
\triangle v_{shear} = \sqrt{ \rm G M r_{\rm cloud}^{2} \over d_{\rm bh}^{3} },
\label{eq5:shear}
\end{equation}

\noindent
where G is the gravitational constant, r$_{\rm cloud}$ is the cloud radius, and M is the total enclosed mass within $\rm d_{bh}$, which is dominated by the black hole, i.e., M $\simeq$ M$_{\rm bh}$. In this equation, $\rm d_{\rm bh} \gg r_{\rm cloud}$. The imposed shear across the model cloud is on the order of 2.2 km/s. This affects the cloud morphologically and adds turbulence to the system. The shearing turbulence is comparable to, but slightly lower than, the initial turbulence of 5 km/s. The shearing time, $\rm t_{shear}$ = 2r$_{\rm cloud}/\triangle \rm v_{shear}$, is about three times larger than the initial free-fall time of the cloud, where $\rm t_{ff}$ is given by

\begin{equation} \rm
t_{ff} = \sqrt{\frac {3\pi} {32{\rm G}\rho}} \simeq 10^{5} \rm \,yr.
\label{eq:freefall}
\end{equation}

\noindent
Thus, a gravitationally bound molecular cloud of 10$^{5}$ cm$^{-3}$ is able to exist in such an environment. As mentioned before, we initiate the molecular cloud with turbulent conditions that are typical for an active environment, FWHM = 5 km/s, or $\rm \sigma_{turb}$ = 2.1 km/s \citep{2009A&A...503..459P}, and apply this over all scales with a power spectrum of $\rm P(k) \propto k^{-4}$, following the empirical laws for compressible fluids \citep{1981MNRAS.194..809L, 1999ApJ...522L.141M, 2004ApJ...615L..45H}. This scaling is also known as Burgers turbulence \citep{Burgers1939, 2007PhR...447....1B}. Besides the injected energy of gravitational shear, we do not drive the turbulence in this work.

For the initial magnetic field, we associate our model cloud with the observed star-forming region RCW 38 surrounded by a molecular cloud \citep{2001ApJ...554..916B, 2008hsf2.book..124W} with a magnetic field strength of B = 38 $\mu$G. We adopt this value and initialize a uniform magnetic field strength throughout the domain. The orientation of the initial magnetic field vectors is chosen to be a uniform field with equal strength along each spatial direction, i.e., the field lines make an angle of 45 degrees with each axis. To study and compare against molecular clouds with higher magnetic field strengths, we compute another model with B = 135 $\mu$G. The higher field strength is comparable to molecular clouds such as NGC 2024 and the infrared dark cloud G28.34+0.06. These molecular clouds have similar density (10$^{5} \rm \,cm^{-3}$) and size (few$\times0.1$ pc) as the model cloud.

\begin{table*}
\begin{center}
\caption{Model parameters}
\begin{tabular}{cccccccc}
\hline
\hline
Model	& B-field strength	& Magnetic flux $\Phi$	& M/$\Phi$	& Magnetic pressure	& $\tau_{\rm ad}$	& X-ray flux	& \\
	& [$\mu$G]		& [Gauss cm$^{-2}$]	& [wrt critical]& [dyne cm$^{-2}$]	& [yr]			& [erg s$^{-1}$ cm$^{-2}$]& \\
\hline
Model A	& 0			& 0			& --		& 0			& 0			& 0	& \\
Model B	& 38			& $1.20\times10^{32}$	& 21.76		& 5.75$\times10^{-11}$	& $5.20\times10^7$	& 0	& \\
Model C	& 135			& $4.24\times10^{32}$	& 6.12		& 7.25$\times10^{-10}$	& $6.57\times10^8$	& 0	& \\
Model D	& 38			& $1.20\times10^{32}$	& 21.76		& 5.75$\times10^{-11}$	& $5.20\times10^7$	& 160	& \\
Model E	& 135			& $4.24\times10^{32}$	& 6.12		& 7.25$\times10^{-10}$	& $6.57\times10^8$	& 160	& \\
\hline
\end{tabular} \\
\hspace{-7.9cm} {\footnotesize Note: The primary cloud models are models A, B, and C.}
    \label{tab:parameters}
\end{center}
\end{table*}

We assume that the magnetic flux is frozen to the gas. Flux-freezing holds even for primordial clouds as long as the magnetic force is smaller than the gravitational force during collapse \citep{2004ApJ...609..467M, 2007PASJ...59..787M}, and should become even better with increasing ionization degree due to cosmic rays or radiative backgrounds. In order to be able to collapse, the ratio of gaseous mass over the magnetic flux needs to be larger than a critical value. \cite{2008A&A...487..247F} derived the mean mass-to-flux ratio $\lambda$ for a sample of 14 clumps of massive star formation using their measured line of sight magnetic field strengths. They found this ratio to be $\rm \lambda \simeq 6 \pm 0.5$, where $\lambda$ is defined as

\begin{equation} \rm
\lambda=\frac{M} {\Phi} / \left( \frac{M} {\Phi} \right)_{crit} = c_{\Phi}^{-1} \sqrt{G} \frac{M} {\Phi}.
\label{eq:masstoflux}
\end{equation}

\noindent
Here, $\Phi$ is the magnetic flux, $\rm (M/\Phi)_{crit}$ is the critical mass-to-flux ratio, and $\rm c_{\Phi}$ \citep{1976ApJ...210..326M} is a factor that is expected to be about $0.12-(2\pi)^{-1}$ \citep{1976ApJ...210..326M, 1978PASJ...30..671N, 1988ApJ...335..239T}. We have adopted the value of $(2\pi)^{-1}$ for $\rm c_{\Phi}$. For the model clouds in this study, given our fixed clump mass of 800 M$_{\odot}$ and field strengths of 38 $\mu$G and 135 $\mu$G, we have mass-to-flux ratios of $\lambda=21.8$ and 6.1, respectively. This means that both model clouds are supercritical ($\lambda > 1$, i.e., prone to gravitational collapse). The model parameters are listed in Table \ref{tab:parameters}.

% \begin{table*}
% \begin{center}
% \caption{Model parameters}
% \begin{tabular}{cccccccc}
% \hline
% \hline
% Model	& B-field strength	& Magnetic flux $\Phi$	& M/$\Phi$	& Magnetic pressure	& $\tau_{\rm ad}$	& X-ray flux	& \\
% 	& [$\mu$G]		& [Gauss cm$^{-2}$]	& [wrt critical]& [dyne cm$^{-2}$]	& [yr]			& [erg s$^{-1}$ cm$^{-2}$]& \\
% \hline
% Model A	& 0			& 0			& --		& 0			& 0			& 0	& \\
% Model B	& 38			& $1.20\times10^{32}$	& 21.76		& 5.75$\times10^{-11}$	& $5.20\times10^7$	& 0	& \\
% Model C	& 135			& $4.24\times10^{32}$	& 6.12		& 7.25$\times10^{-10}$	& $6.57\times10^8$	& 0	& \\
% Model D	& 38			& $1.20\times10^{32}$	& 21.76		& 5.75$\times10^{-11}$	& $5.20\times10^7$	& 160	& \\
% Model E	& 135			& $4.24\times10^{32}$	& 6.12		& 7.25$\times10^{-10}$	& $6.57\times10^8$	& 160	& \\
% \hline
% \end{tabular}
%     \label{tab:parameters}
% \end{center}
% \end{table*}

Considering that the magnetic pressure is $\rm P_{mag} = B^{2}/8\pi$, the initial pressure exerted by the Lorentz force inside the cloud becomes $\rm P_{mag} = 5.75 \times 10^{-11}$ dyne cm$^{-2}$ for model B (38 $\mu$G) and $\rm 7.25 \times 10^{-10}$ dyne cm$^{-2}$ for model C (135 $\mu$G). Since the thermal pressure for an isothermal temperature of 10 K is $\rm P_{th}$ = $1.38 \times 10^{-10}$ dyne cm$^{-2}$, these pressures yield a $\rm \beta = P_{th}/P_{mag} = 2.40$ for model B and $\rm \beta = 0.19$ for model C.

\begin{figure}[htb!]
\centering
\includegraphics[scale=0.51]{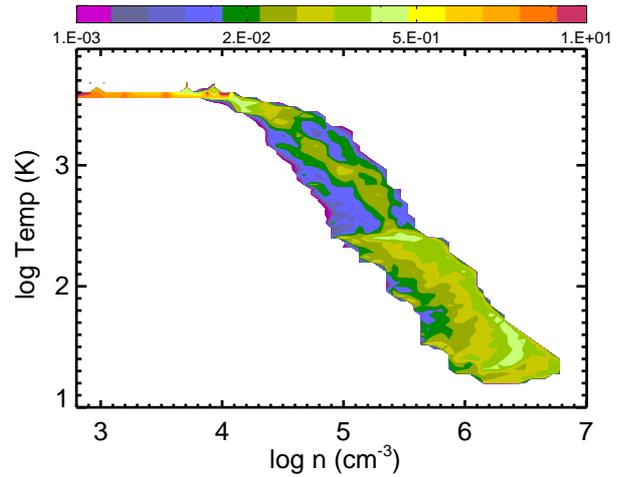}
\caption{Temperature-density phase diagram of an X-ray irradiated simulation (model D) at $\rm t=2 \,t_{ff}$. The color in this figure represents the amount of mass (in M$_{\odot}$) lying in a contoured region.}
\label{fig:phase}
\end{figure}

The temperature of the primary cloud models remain isothermal at 10 K. The isothermal sound speed of the cloud is $\rm c_{s}=0.19$ km s$^{-1}$ in these runs. We also run the simulations B and C for the case of an active black hole and name them models D and E. An active black hole will produce strong X-rays ($\rm 1-100$ KeV) in its accretion disk. A 10$^{7}$ M$_{\odot}$ black hole accreting at 10\% Eddington radiates with a bolometric flux of $\sim$10$^{4}$ erg s$^{-1}$ cm$^{-2}$ at 10 pc distance. Most (90\%) of the energy is emitted at optical and UV wavelengths, but these wavelengths will be attenuated along the line of sight by large columns of gas and dust. Such environments are typical of obscured AGN, as formed in (U)LIRGS, with obscuring columns of 10$^{22}$-10$^{23.5}$ cm$^{-2}$ \citep{2005Ap&SS.295..143A, 2007A&A...476..177P, 2008A&A...488L...5L}. X-rays, however, can penetrate large columns of gas and dust and dominate the thermal balance up to distances of 300 pc from the black hole \citep{2010A&A...513A...7S}. The expected ($\rm 1-100$ KeV) X-ray flux at 10 pc from the black hole is $\sim$160 erg s$^{-1}$ cm$^{-2}$ \citep{2010A&A...522A..24H}. The cloud models with X-rays obtain their temperatures from the XDR code, which can go up to $\sim 5 \times 10^{3}$ K \citep{2010A&A...522A..24H}. Fig. \ref{fig:phase} displays the temperature-density phase diagram obtained from one of the X-ray models. 
Under such conditions, the sound speed will rise to a maximum of 5 km/s in the hottest parts of the X-ray irradiated cloud. A temperature of a few$\times 10^{3}$ K results in a thermal pressure of $\rm P_{th} \simeq 8 \times 10^{-8}$ dyne cm$^{-2}$, which is $2-3$ orders of magnitude higher than the magnetic pressure, $\beta \simeq 110-1400$.

Ambipolar diffusion according to the conditions of our model is not expected to play an important role for our calculations, because it operates on a much longer timescale than the free-fall time of $\rm t_{ff} = 10^{5}$ yr. The ambipolar diffusion timescale

\begin{equation} \rm
\tau_{ad} = 3\times10^{6} ~yr \,\left( \frac{n} {10^{4} \,cm^{-3}} \right)^{3/2} \left( \frac{B} {30 \,\mu G} \right)^{-2} \left( \frac{r_{cloud}} {0.1 \,pc} \right)^{2}
\label{eq:tauad}
\end{equation}

\noindent
is on the order of $\rm \tau_{ad}$ = $5.2-65.7\times10^{7}$ yr for the two magnetized cloud models with magnetic field strengths of 38 and 135 $\mu$G. The ambipolar diffusion can work in clouds with lower ionization degree or at smaller scales like compact cores inside the cloud. In such a small area, $\rm r_{cloud}$ is much smaller and the magnetic field is much stronger (but the density will be larger). According to \cite{2005ApJ...631..411N}, the ambipolar diffusion can be accelerated by the turbulent compression and therefore may play a role in the evolution of cores out of which stars form, in particular in turbulent clouds. For our X-ray models, the ionization degree will be higher which will make ambipolar diffusion less efficient.

\section{The collapsing cloud}
\label{sec:collapsingcloud}
\subsection{Cloud morphology}
\begin{figure*}[htb!]
\begin{tabular}{ccc}
\begin{minipage}{5.5cm}
\includegraphics[scale=0.36]{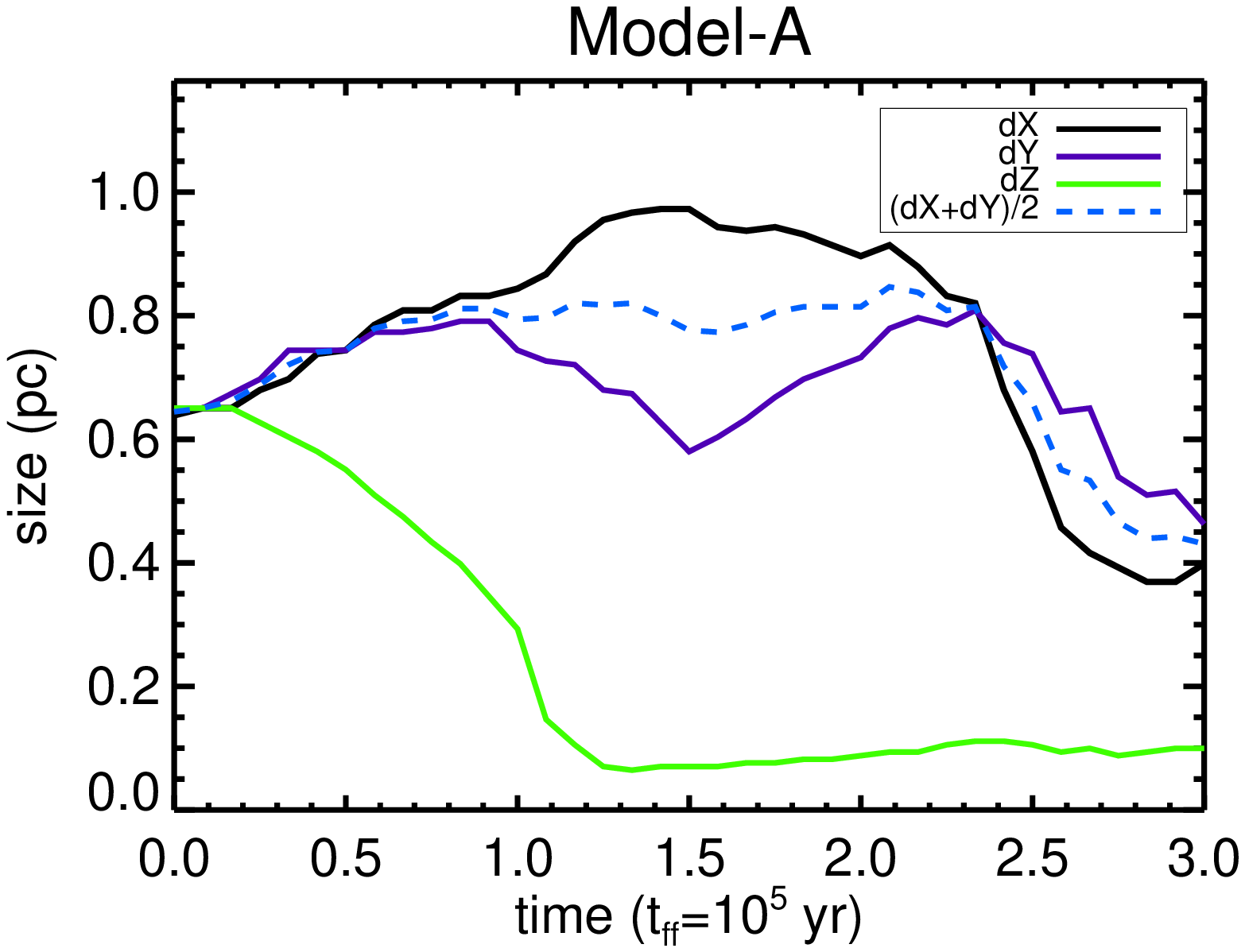}
\end{minipage} &
\begin{minipage}{5.5cm}
\includegraphics[scale=0.36]{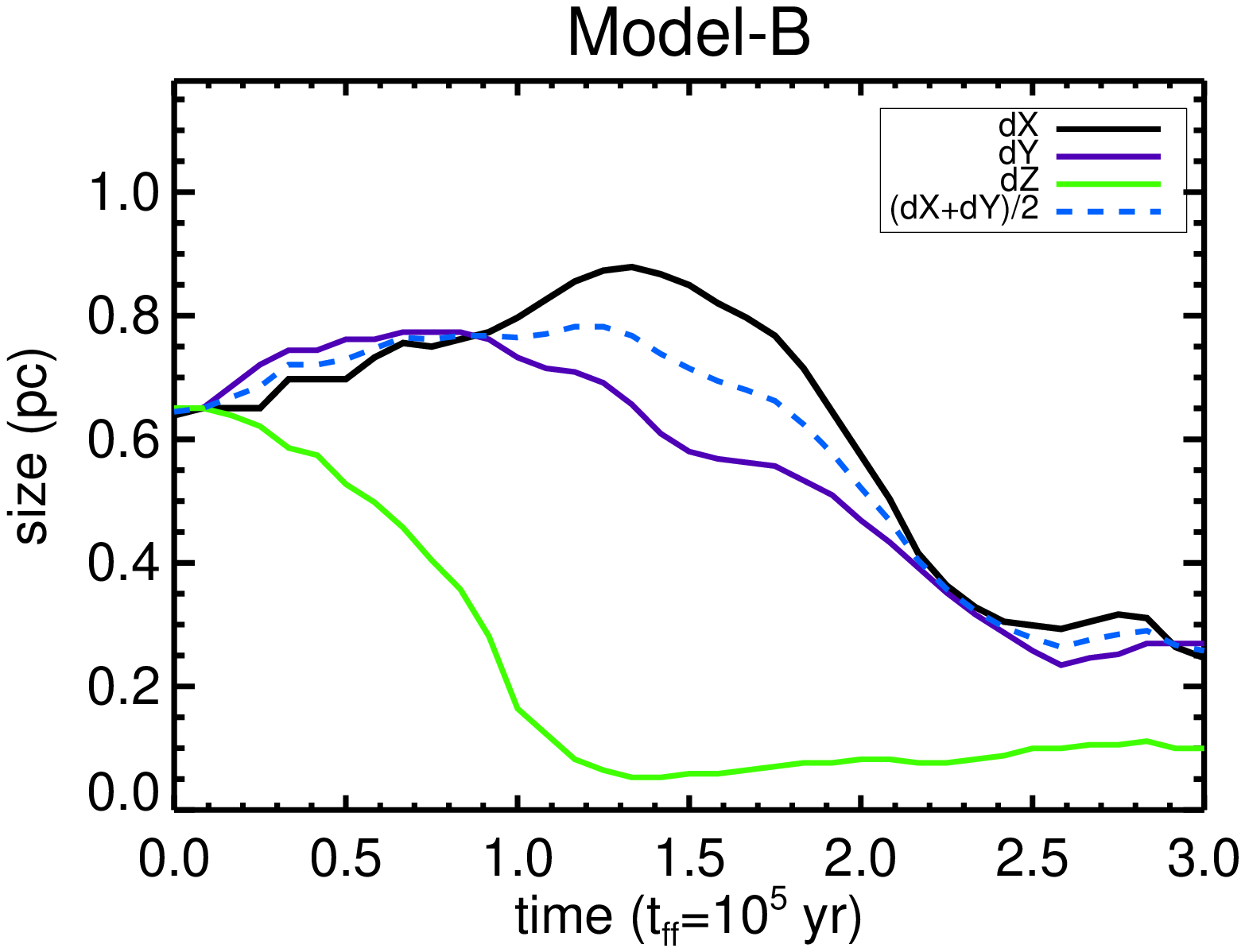}
\end{minipage} &
\begin{minipage}{5.5cm}
\includegraphics[scale=0.36]{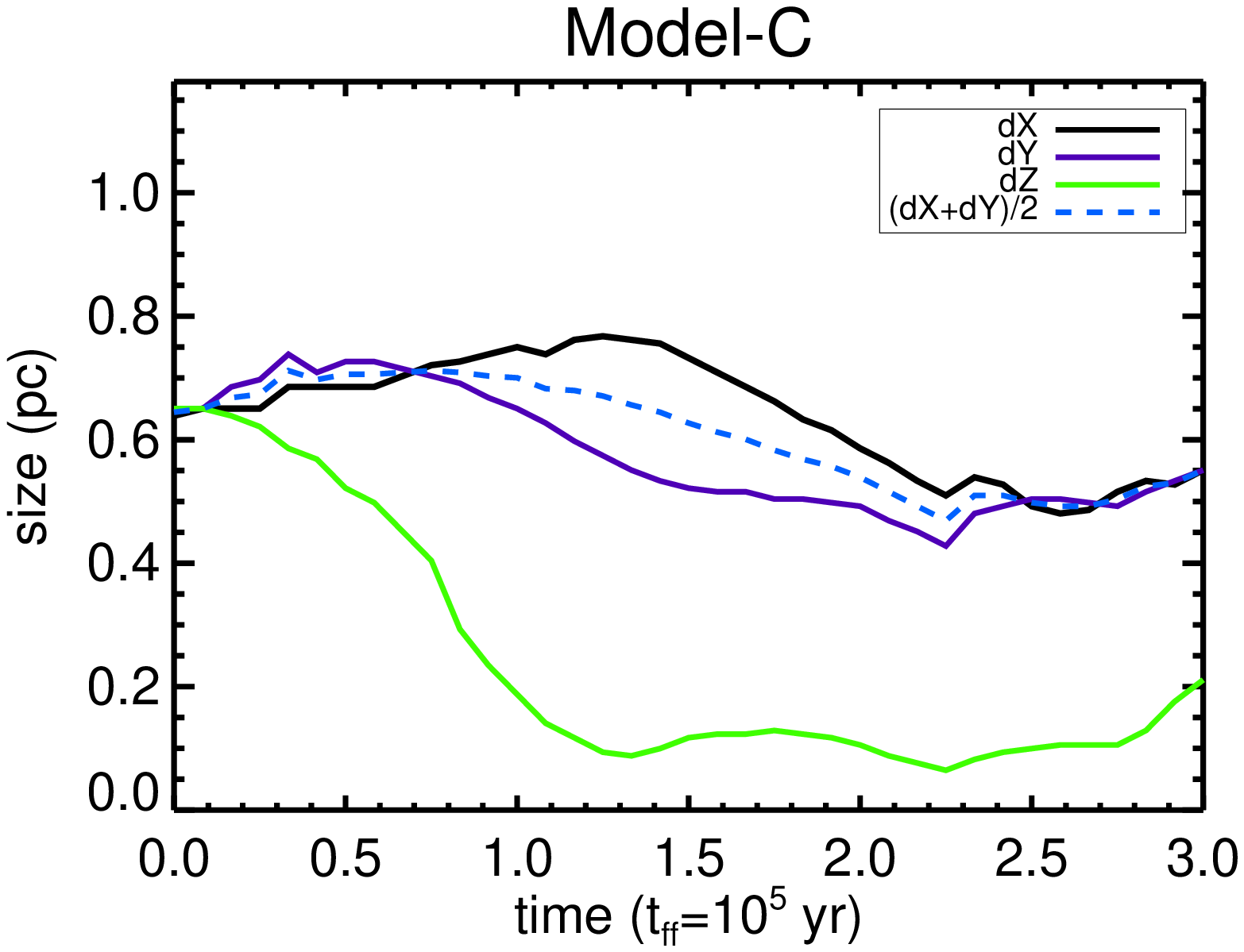}
\end{minipage} \\
\end{tabular}
\caption{Spatial sizes of the cloud. The size of the cloud in each spatial direction is plotted against time in $\rm t_{ff}$. The sizes in the direction of rotation are denoted by dX and dY, and the height is given by dZ. The dashed line is a representation of the average disk size.}
\label{fig:shape}
\end{figure*}

A molecular cloud that is gravitationally unstable will collapse in about one free-fall time, which is about $10^{5}$ years in all our simulations. By virtue of geometrical effects for an object in orbit around a gravitating body and due to the tidal shear imposed by the gravity of the black hole, any symmetrical collapse is disrupted. Clouds in a strong gravity field will collapse in one direction first and quickly form a disk. Furthermore, the existing gravitational shear generates turbulence and pressure in the two spatial directions of the orbital plane, while the third spatial direction (Z-direction) experiences no additional pressure and can proceed with a near free-fall collapse. In Fig. \ref{fig:shape}, we plot the size of the cloud from the X, Y, and Z perspective against time. 
From the cloud sizes in this figure, we can see that a disk forms in the plane of the orbit within one free-fall time. The increase in density due to this asymmetric collapse creates even more pressure along the orbital plane of the cloud. In the same figure we also see that this pressure causes a slight expansion of the cloud in the XY-plane.

The cloud itself is selected by taking the gas densities above $\rm 10^{5} \,cm^{-3}$ and the sizes are selected for densities above the mean cloud density. The collapse in the Z-direction continues unimpeded for each model, but slows down between $\rm 1-1.5 \,t_{ff}$. The disk thickness stabilizes around 0.1 pc in all runs. We can see that the shape of the cloud in the X and Y directions stays roughly the same. The expansion of the disk, which can be inferred from the dashed line ($\rm (dX+dY)/2$), is explained by the gravitational shear stretching the cloud along the orbital plane (the XY-plane) in the direction of rotation and by the collapse in the Z-direction which increases the pressure in the same plane. We see that disk formation is not affected by the presence of magnetic fields of the chosen magnitudes in this work.

\begin{figure}[htb!]
\centering
\includegraphics[scale=0.16]{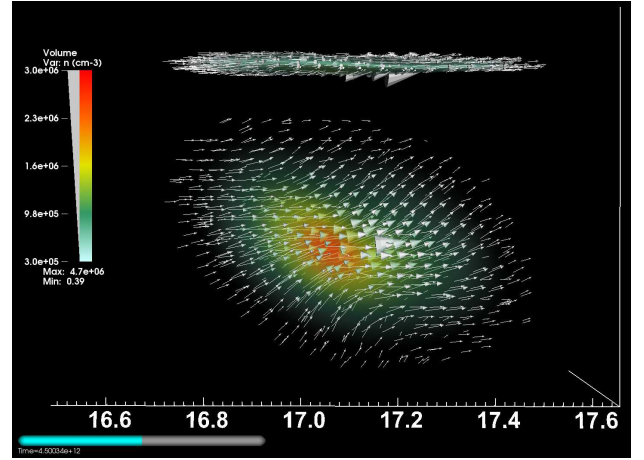}
\caption{Magnetic field vectors versus disk orientation. The magnetic field lines align with the disk between $\rm \sim1.0-1.5 \,t_{ff}$. The image shows a volume plot of the density from an XZ perspective (top) and an XY perspective (bottom) at $\rm t=1.5t_{ff}$. The magnetic field vectors are displayed as white arrows, where the field strength B typically ranges from $\rm 1 - 1000 \,\mu G$. The X-axis is given at the bottom in units of parsec.}
\label{fig:fieldlines}
\end{figure}

The magnetic field lines align parallel to the disk in less than a free-fall time as the disk forms. The alignment of the magnetic field with the disk is because the gravity is more important than the magnetic force. The vector components along the XY-plane get enhanced by the increase in density in the course of gravitational collapse. Moreover, because of tidal shear from the black hole, magnetic field components parallel to the disk will form rather quickly and only the parallel components are amplified when the field is compressed by the flow. The field lines are completely aligned with the disk around $\rm 1-1.5 \,t_{ff}$, while, at the same time, the field lines in the orbital plane shape like an hourglass. An image to this effect is shown in Fig. \ref{fig:fieldlines}.

\subsection{Turbulence}
\begin{figure*}[htb!]
\begin{tabular}{ccc}
\begin{minipage}{5.5cm}
\includegraphics[scale=0.36]{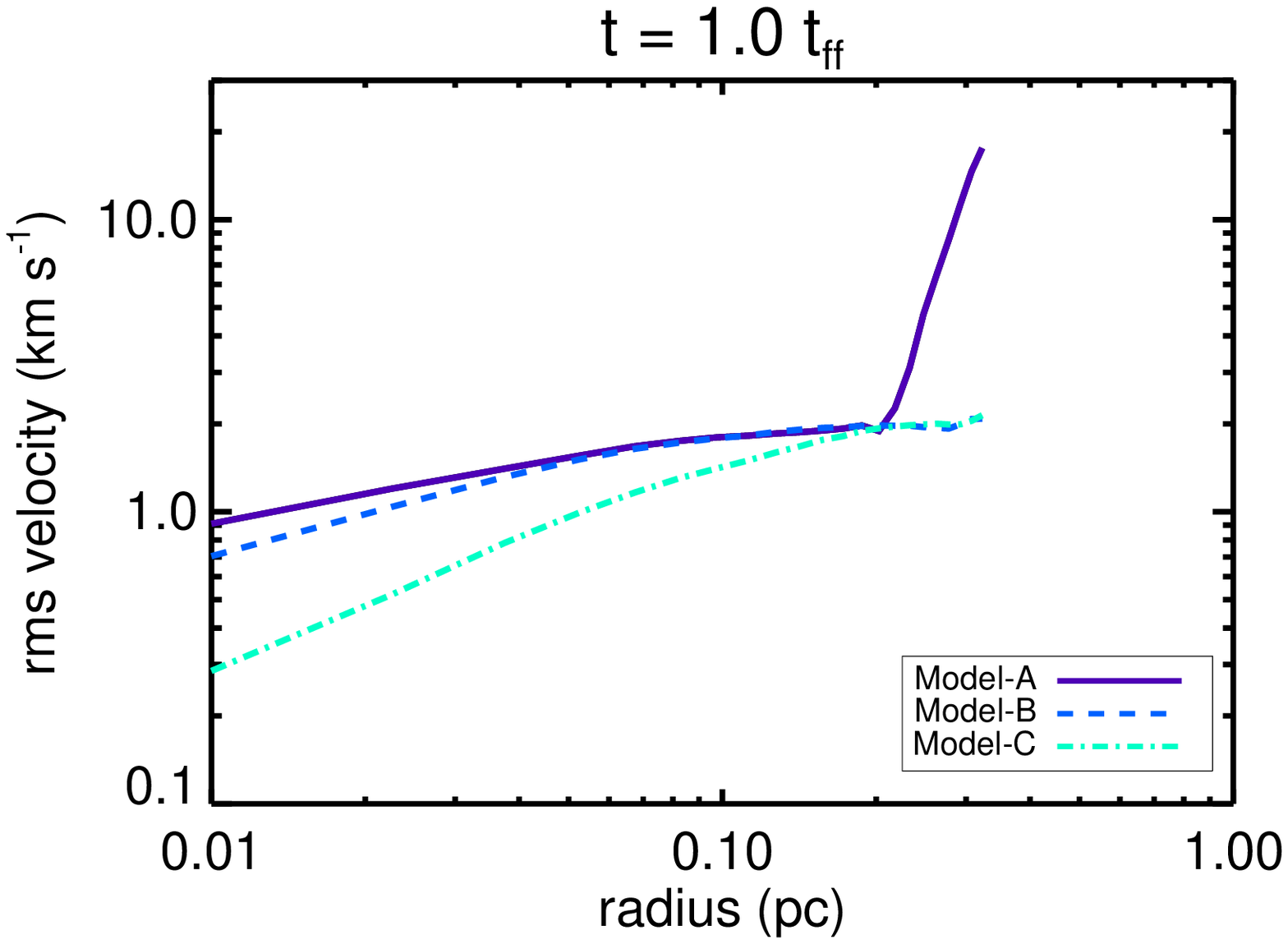}
\end{minipage} &
\begin{minipage}{5.5cm}
\includegraphics[scale=0.36]{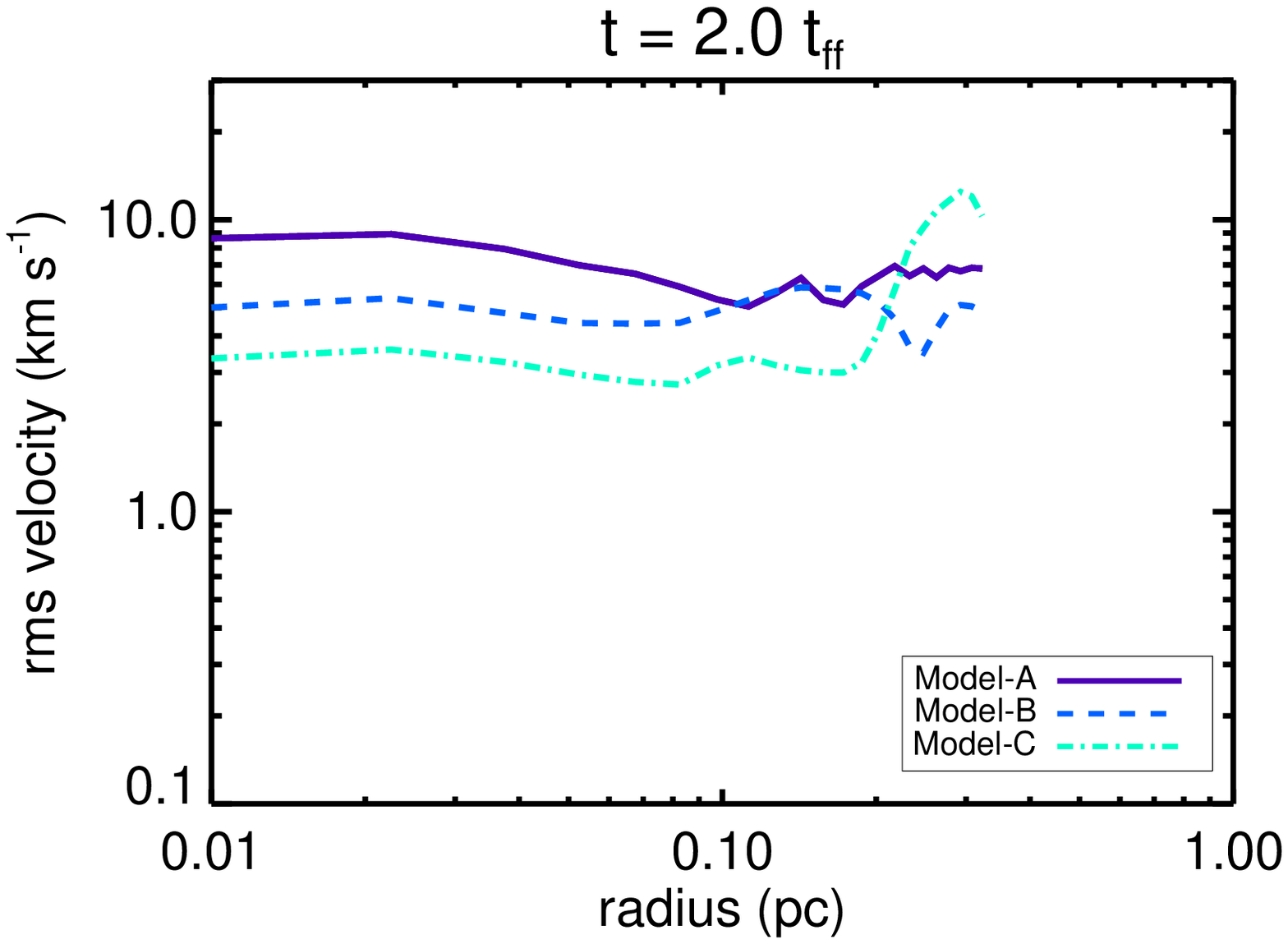}
\end{minipage} &
\begin{minipage}{5.5cm}
\includegraphics[scale=0.36]{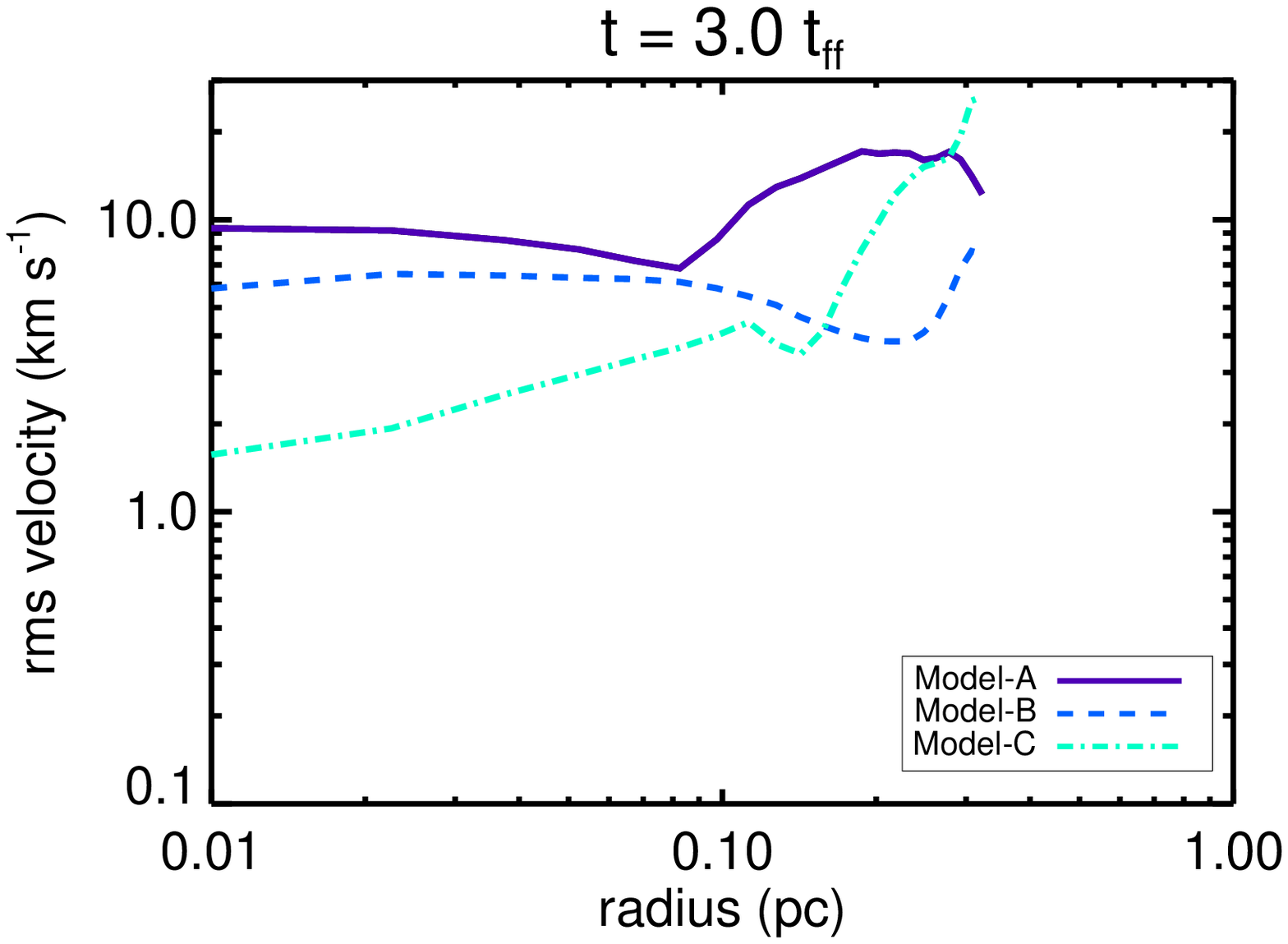}
\end{minipage} \\

\begin{minipage}{5.5cm}
\includegraphics[scale=0.36]{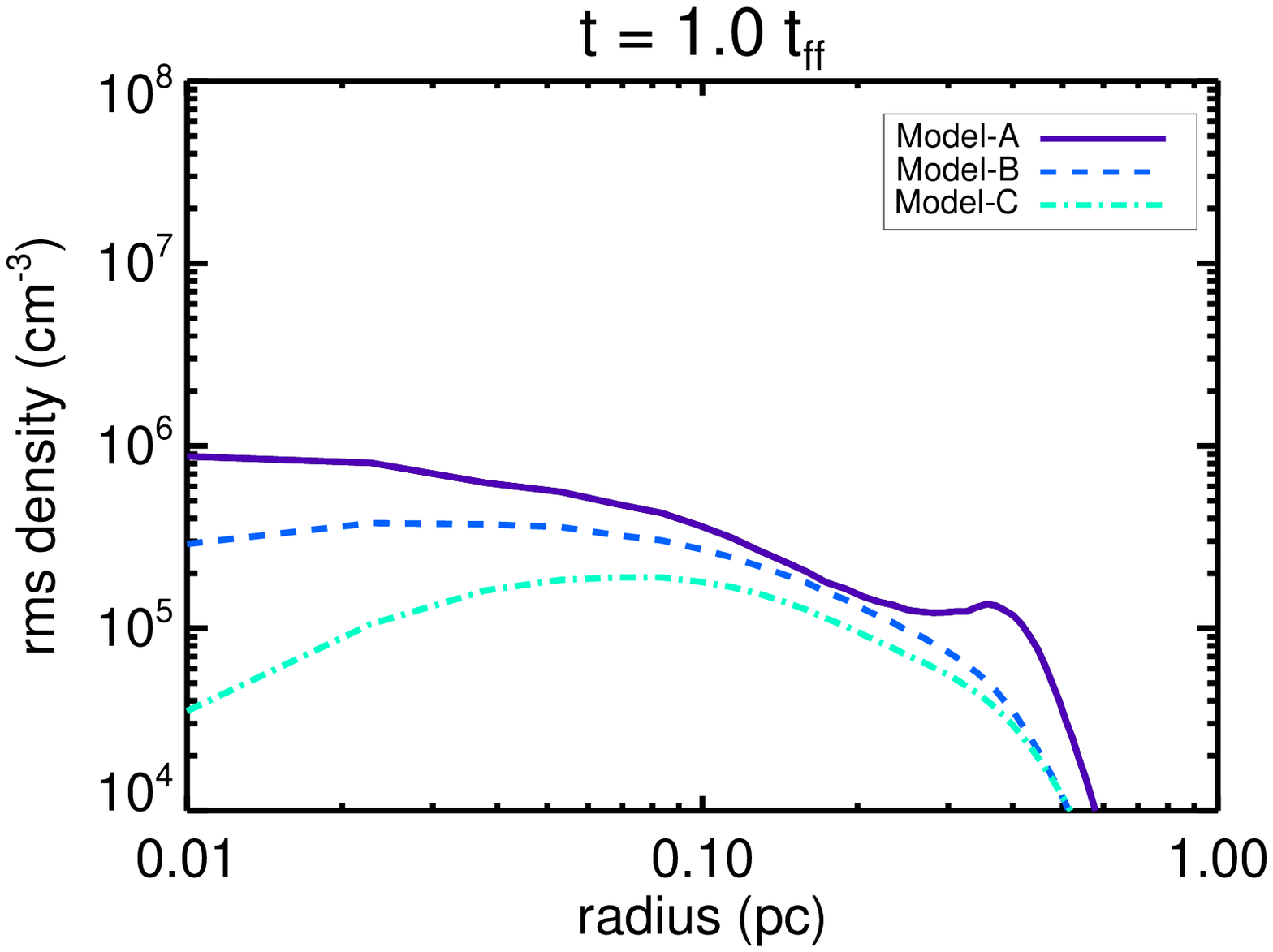}
\end{minipage} &
\begin{minipage}{5.5cm}
\includegraphics[scale=0.36]{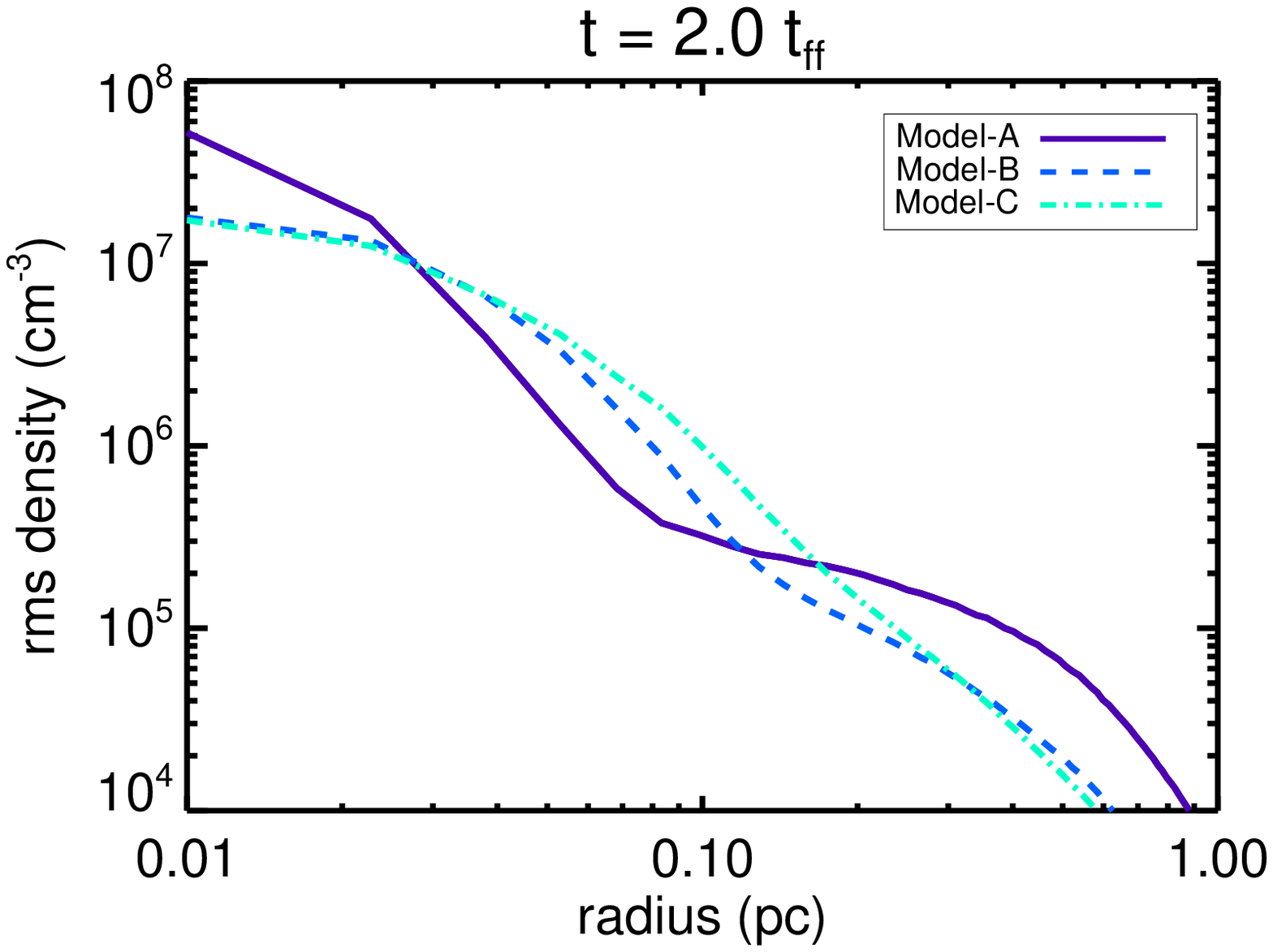}
\end{minipage} &
\begin{minipage}{5.5cm}
\includegraphics[scale=0.36]{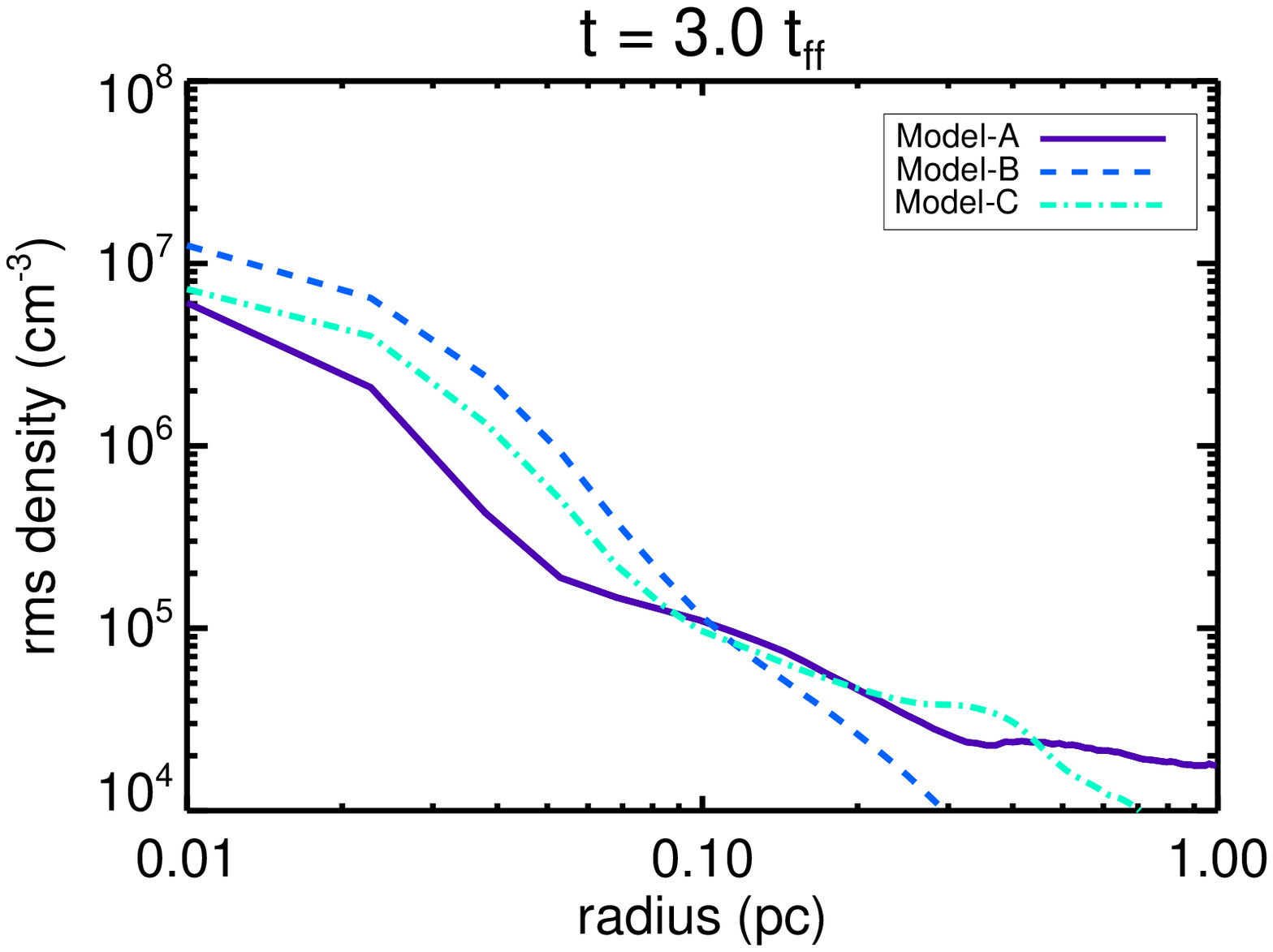}
\end{minipage}
\end{tabular}
\caption{Radial profiles of the cloud. The rms velocities and the rms densities of models A, B, and C are plotted against cloud radius. The radius originates in the center of the cloud. The rms velocities are obtained in the frame of reference of the cloud orbiting the black hole. From left to right, the time frames of t = 1, 2, and 3 $\rm t_{ff}$ are displayed.}
\label{fig:rms}
\end{figure*}

The model cloud develops turbulence during its evolution as a result of the strong gravity field of the black hole. To demonstrate this, we show radial profiles of the rms velocity and the rms density of the cloud for three different time frames in Fig. \ref{fig:rms}.
All models start with a cloud that is turbulent with a Mach number of $\mathcal{M}=9$ and a number density of 10$^{5}$ cm$^{-3}$. From the changing velocity field in the different snapshots we can see that the cloud gets more turbulent in time with Mach numbers that grow to around $\mathcal{M}=50$. This means that the turbulence increases by a factor of $5-6$ within the time frame of $\rm 3 \,t_{ff}$.

We also notice that the rms velocities are initially lower at small radii and higher at large radii. This can be inferred from the rms velocities displayed in the top left panel of Fig. \ref{fig:rms}. Since we take the mean of the velocity at each radius to compute the rms, it is appropriate to state that the radius is analogous to the scale length. At $\rm t = 2 \,t_{ff}$, the turbulence at smaller scales is increased by the decay of turbulence from larger scales.

\begin{figure}[htb!]
\centering
\includegraphics[scale=0.36]{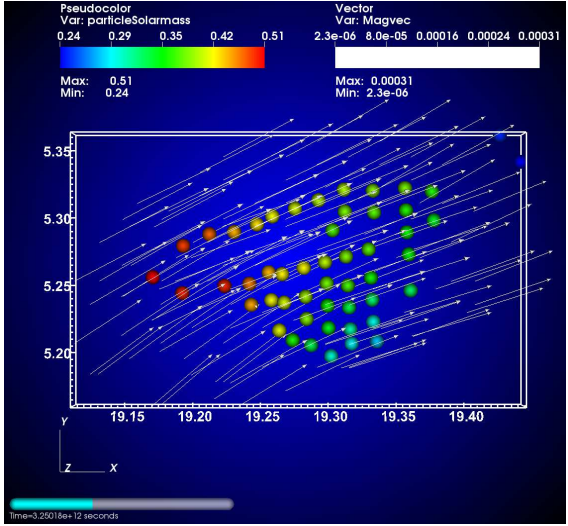}
\caption{Star formation along the field lines. Sink particles (colored spheres) are aligned with the magnetic field (white arrows) for model B at $\rm t=1.1t_{ff}$, i.e., just after the first sinks have formed. The masses of sink particles are given in solar mass units and the magnetic field strength is given in Gauss. The axes are given in parsec.}
\label{fig:starfield}
\end{figure}

We find that the rms velocities are lower with increasing magnetic field strength. This is especially true at smaller radii. Given that the density is the highest close to the center of the cloud, the magnetic field strength is also greater there. Additionally, the rms densities are also smaller with magnetic fields. These lower rms values indicate that there is less cloud fragmentation with increasing magnetic field strength. In all cases, the mean density is also lower at the same moment in time with increasing magnetic field strength. From the reduced density build-up, we can understand that magnetic fields successfully slow down collapse. However, the difference between the three models in the rms density and the mean density decreases after $\rm t = 1.5 \,t_{ff}$.

\section{Results}
\label{sec:results}
\subsection{Star formation along magnetic field lines}
We see that most of the sink particles form along the field lines and, especially, where the field lines are the strongest, which is generally the central part of the cloud. The first sink particles are created at $\rm t \simeq 1 \,t_{ff}$ around densities of $\rm n \geq 10^{6}$ cm$^{-3}$, with a thermal Jeans mass of $\rm M_{J} \leq 0.5 \,M_{\odot}$. The magnetic field strength has increased by a factor of $10-12$ from the initial value at the onset of star formation. Fig. \ref{fig:starfield} illustrates the sink particle formation along the magnetic field lines just after the first sink particles have formed.
%
%
% \begin{figure}[htb!]
% \centering
% \includegraphics[scale=0.37]{Sink-n-bfield-new1.ps}
% \caption{Star formation along the field lines. Sink particles (colored spheres) are aligned with the magnetic field (white arrows) for model B at $\rm t=1.1t_{ff}$, i.e., just after the first sinks have formed. The masses of sink particles are given in solar mass units and the magnetic field strength is given in Gauss. The axes are given in parsec.}
% \label{fig:starfield}
% \end{figure}
%
%
These results are in accordance with earlier findings \citep[e.g., ][]{2008ApJ...687..354N, 2010ApJ...720L..26L}. The reason why star formation is occurring along the field lines is explained by the directionality of the Lorentz force. The magnetic field increases the plasma pressure, by an amount $\rm B^2/8\pi$, in directions perpendicular to the magnetic field, and decreases the plasma pressure, by the same amount, in the parallel direction. Thus, the magnetic field gives rise to a magnetic pressure, $\rm B^2/8\pi$, acting perpendicular to field lines, and a magnetic tension, $\rm B^2/8\pi$, acting along field lines. This allows the gas density to build up along the field lines until it becomes gravitationally unstable.

\subsection{Star-formation efficiencies}
\begin{figure*}[htb!]
\begin{tabular}{ccc}
\begin{minipage}{5.5cm}
\includegraphics[scale=0.34]{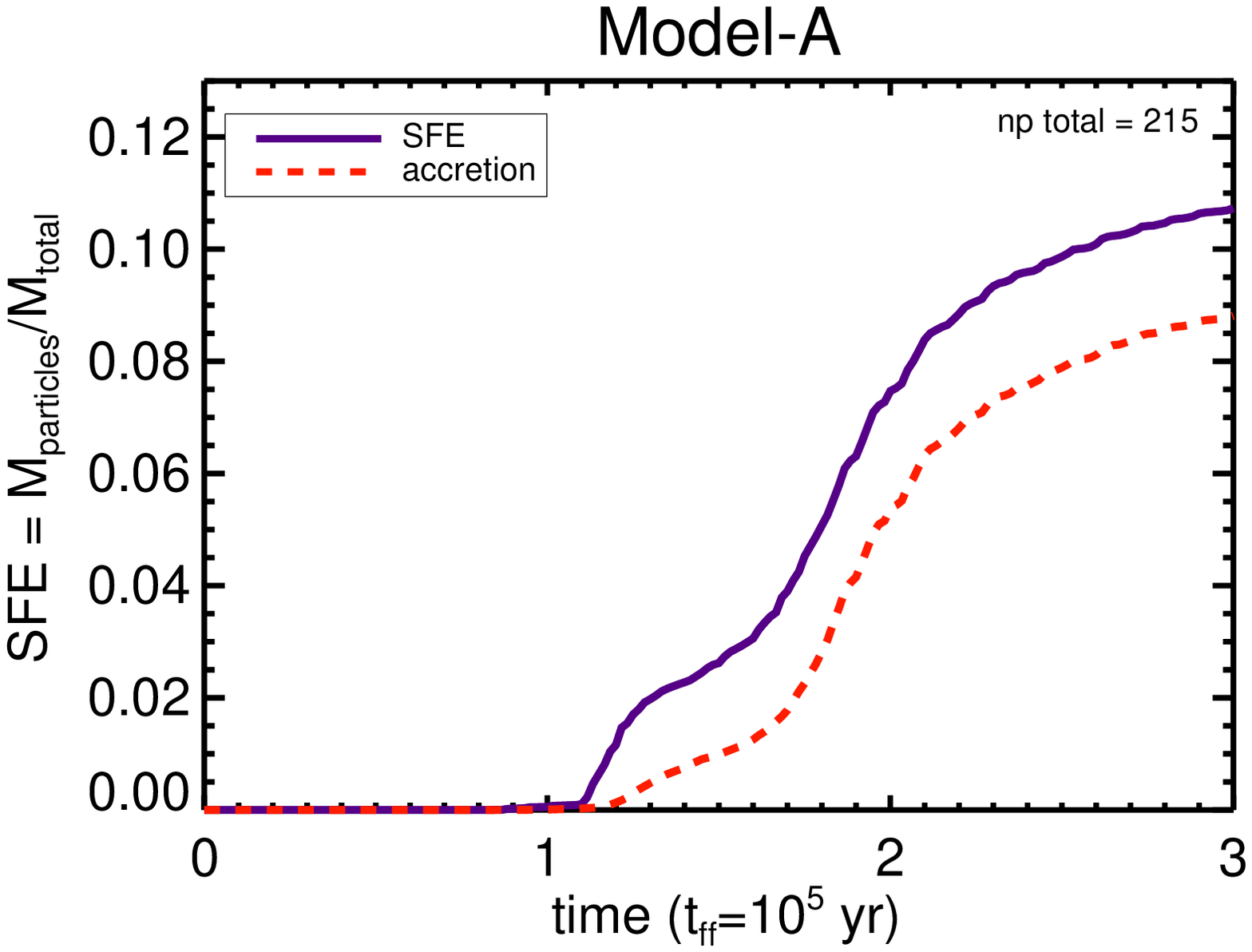}
\end{minipage} &
\begin{minipage}{5.5cm}
\includegraphics[scale=0.34]{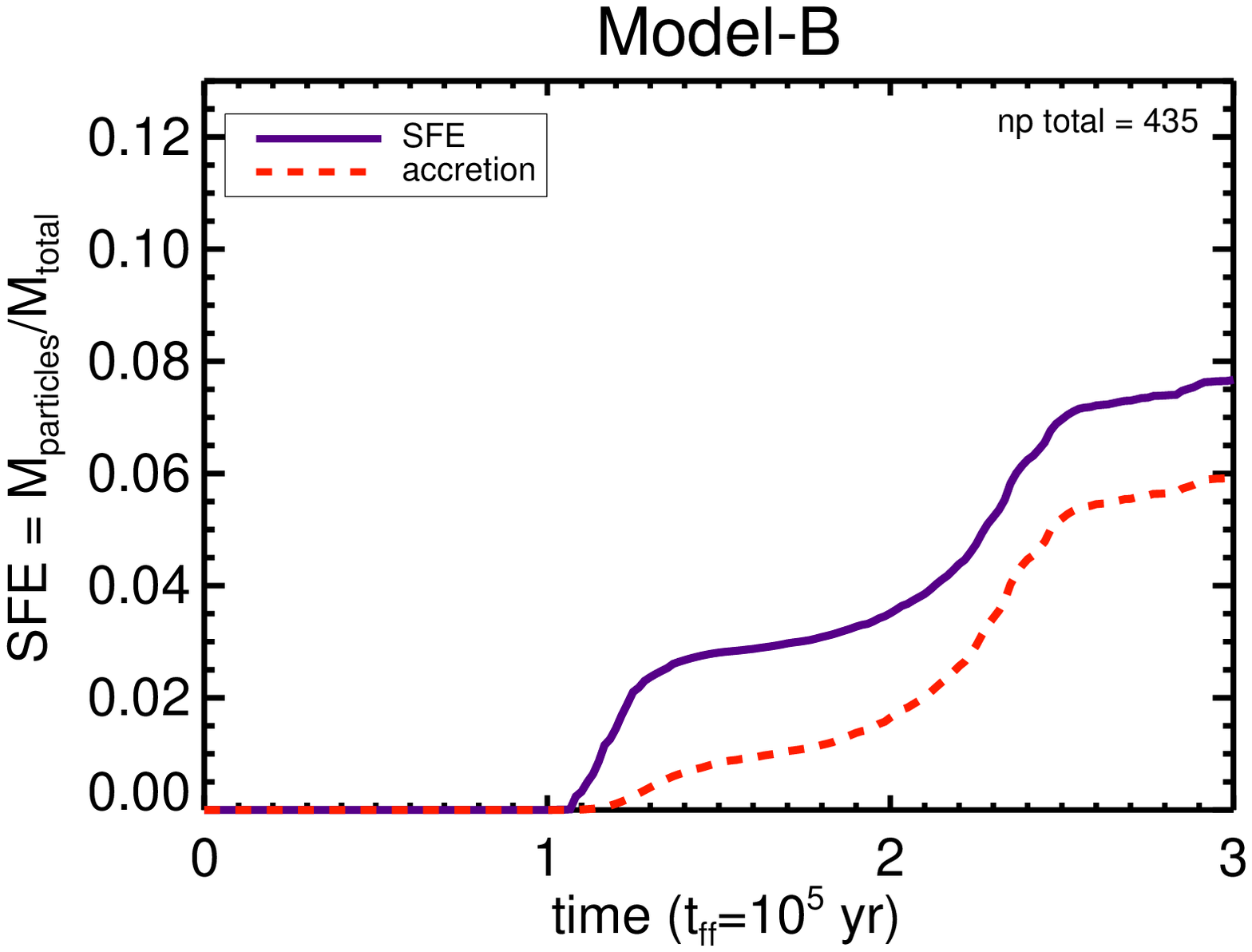}
\end{minipage} &
\begin{minipage}{5.5cm}
\includegraphics[scale=0.34]{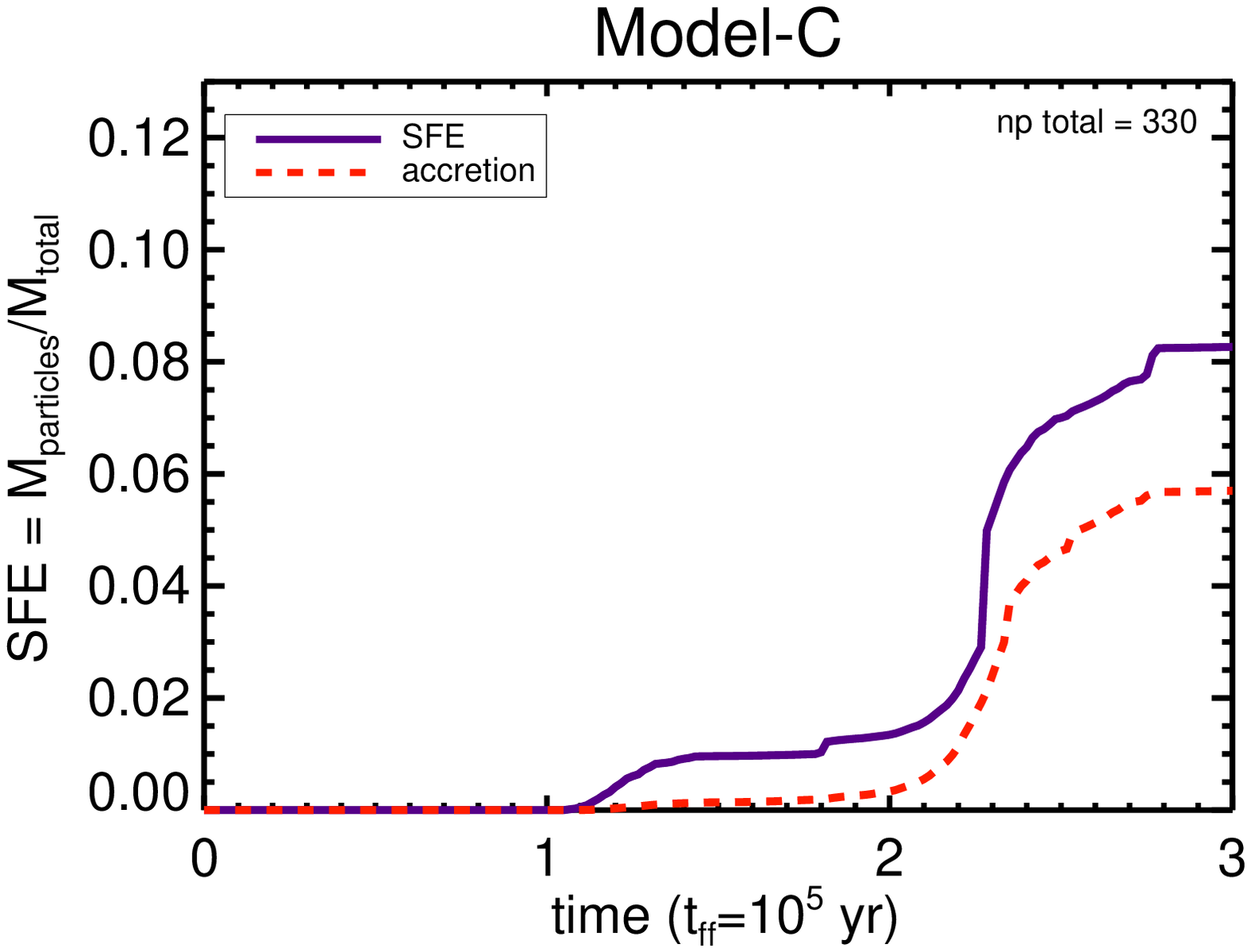}
\end{minipage}
\end{tabular}
\caption{SFEs of the primary models. The ratio of the total sink particle mass over the total mass is plotted against time in $\rm t_{ff}$ as the solid line. The red dashed line portrays the total accreted mass in relationship to the total mass. In the frames from left to right, the initial magnetic field strength increases from 0, 38, to 135 $\mu$G. The total number of sink particles formed during the run is given in the upper right corner of each panel.}
\label{fig:SFEs}
\end{figure*}

{\it Sink particle formation:} 
We display the SFEs of the primary models in Fig. \ref{fig:SFEs}. 
The two lines in each panel of this figure sketch the time evolution of the total sink particle mass (solid line) and the total amount of accreted mass onto sink particles (dashed line) with respect to the total gas mass. Examining the SFE of model A, we see that it follows a typical behavior that is also seen in previous studies \citep{2008MNRAS.386....3C, 2011A&A...536A..41H}. The onset of star formation for model A is at $\rm t=0.85 \,t_{ff}$ and the SFE at $\rm t=3 \,t_{ff}$ is 10.7\%. The final SFEs for models B and C are, on the other hand, somewhat lower. Both magnetic field runs enjoy a final SFE of $\sim$8\%. Typical sink particle masses at the time of their creation is around $0.3-0.8$ M$_{\odot}$ for model A, $0.2-0.5$ M$_{\odot}$ for model B, and $0.2-0.3$ M$_{\odot}$ for model C. In model A, we obtain a total number of sink particles, denoted as np, of 215 as measured at $\rm t = 3 \,t_{ff}$. In the presence of magnetic fields, the number of sink particles increases drastically by a factor of $1.5-2$, e.g. $\rm np = 435$ for model B. We presume this increase to reflect the changes induced in the global dynamics, in particular with respect to cloud morphology, gas reservoir, and turbulence. When going to an even larger field strength (model C), the number of sink particles reduces to $330$, due to the decreased SFE and the effect of magnetic pressure. \\

\noindent
{\it Sink particle accretion:} 
The relatively uniform initial conditions in the cloud prompt it to form the bulk of its sink particles in about a free-fall time. The sink particles that have already formed disrupt the formation of new sink particles in their vicinity by accreting the material around them and by disturbing the general process of sink particle formation \citep{2011A&A...536A..41H}. Because of this, star formation is virtually terminated between $\rm 1.3-1.7 \,t_{ff}$.

We note that the SFE curves flatten out just after the onset of star formation around 1.3 $\rm t_{ff}$ and pick up again above 2 $\rm t_{ff}$. The delay that the magnetic fields cause to star formation clearly has a strong effect on the star-formation history. Magnetic fields reduce the gas flow rates into the center of the cloud by magnetic pressure and also lower the initial stellar masses at their creation. The lower initial masses as well as the reduced gas supply both lead to reduced accretion rates in models B and C. We also see this in the SFEs, which is always smaller in the presence of magnetic fields. As accretion scales with the square of the particle mass, having twice more particles with twice less mass means that the total accretion onto protostars will be reduced by a factor of 2, thus reducing the typical mass scale of the stars. We can see from Fig. \ref{fig:SFEs} that the amount of accreted mass is smaller for magnetic field runs in the first 10$^5$ years after the onset of star formation. On the other hand, this also means that the remaining gas reservoir will maintain a high mass for a longer period. Around $\rm t=2\,t_{ff}$, we see a jump in the SFEs in the magnetic field runs, which may suggest that there are regions in the cloud that collapse at different epochs. Rather than a second burst of star formation, we find that this is mainly due to accretion. Sink particles start to accrete more mass in the magnetic field runs because of the higher remaining density (see Fig. \ref{fig:rms}), but also due to mergers happening in the center of the cloud, where most of the stars lie. Mergers become more likely when the masses of stars are higher. Since the newly merged stars from this epoch are more massive, they tend to accrete more gas as well.

\subsection{The initial mass functions}
\begin{figure*}[htb!]
\begin{tabular}{ccc}
\begin{minipage}{5.5cm}
\includegraphics[scale=0.36]{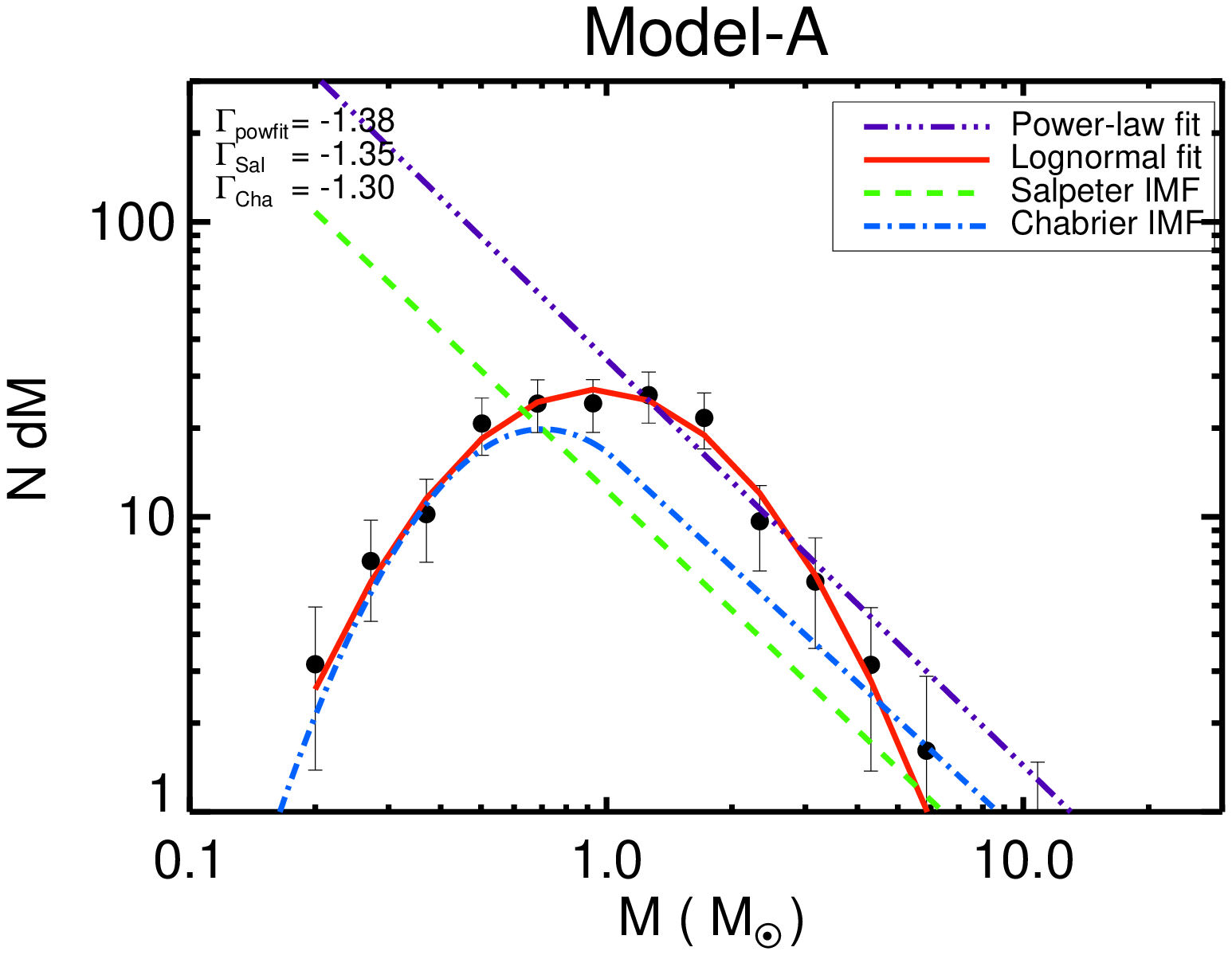}
\end{minipage} &
\begin{minipage}{5.5cm}
\includegraphics[scale=0.36]{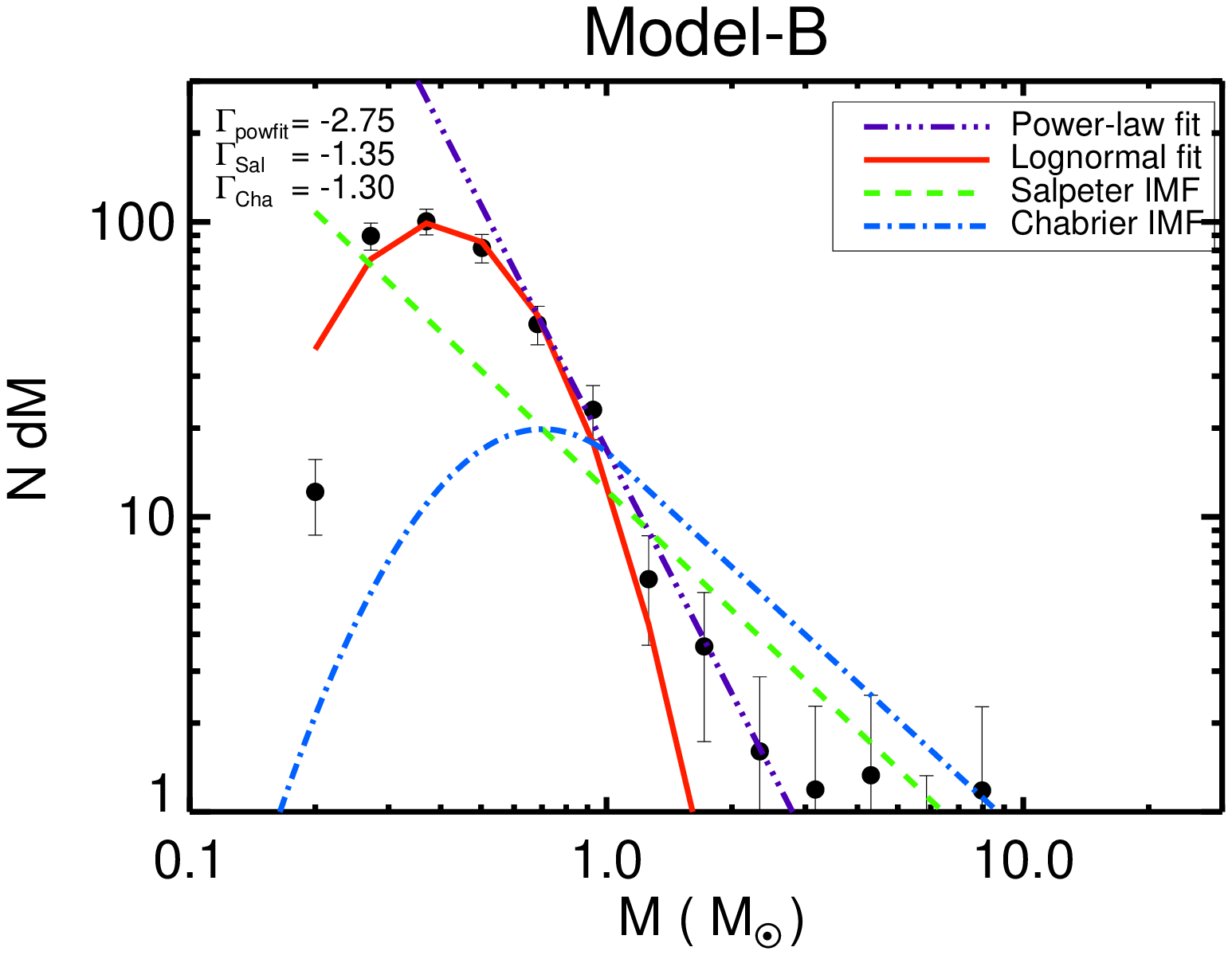}
\end{minipage} &
\begin{minipage}{5.5cm}
\includegraphics[scale=0.36]{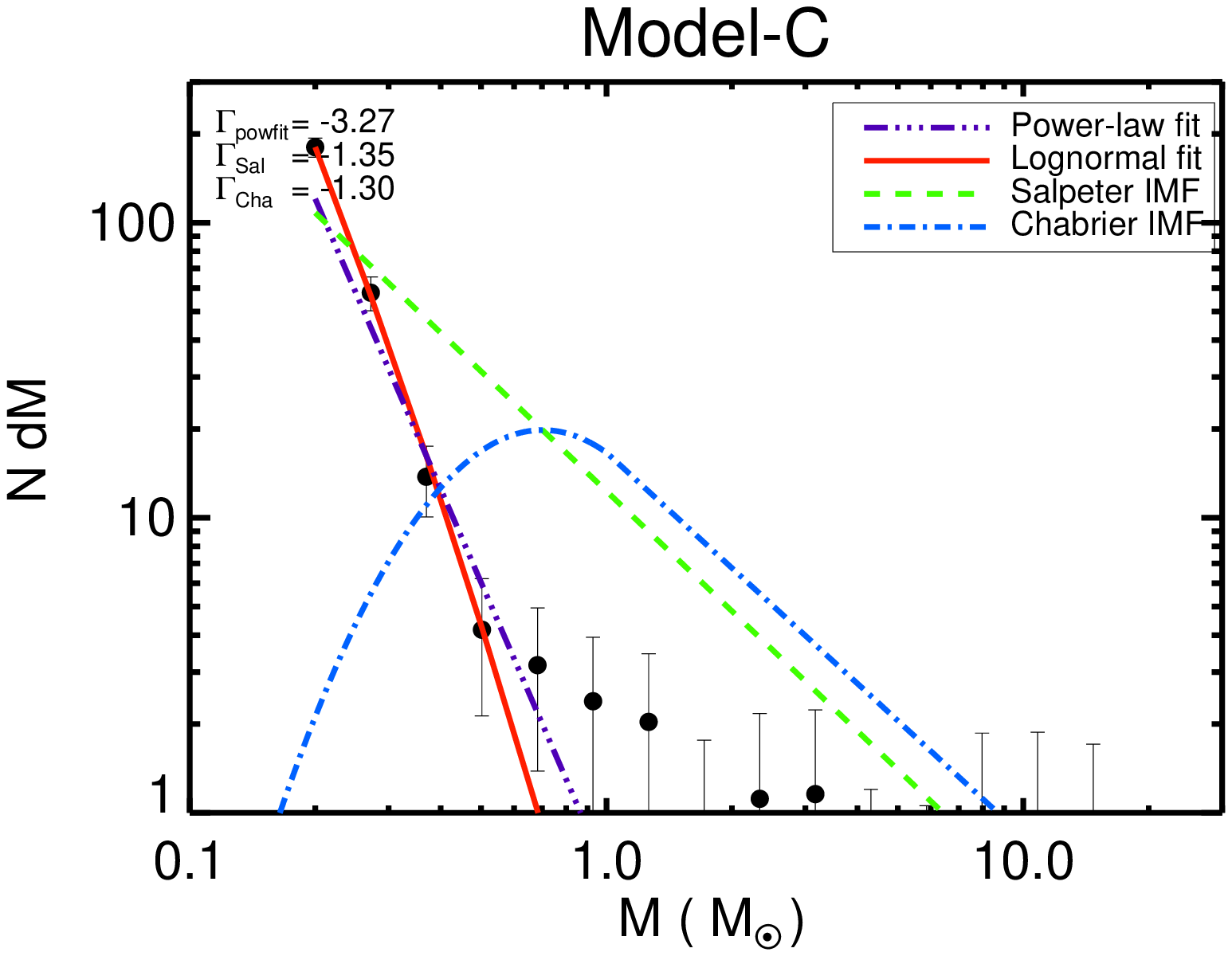}
\end{minipage}
\end{tabular}
\caption{The IMFs of the primary models. The images display the time-averaged IMFs between 1 and 3 free-fall times, where $\rm t_{ff}=10^{5} \,yr$. From left to right, the magnetic field strength increases from 0, 38 to 135 $\mu$G. In each image, for comparison purposes, the Salpeter IMF (green dashed) and the Chabrier IMF (blue dot-dashed) are displayed (with arbitrary values along the Y-axis). Two best fits are applied to the data, a power-law fit and a lognormal fit, and are shown as purple and red lines. With the exception of the lognormal fit, the slopes above the characteristic mass are given in the upper left corner.}
\label{fig:IMFs}
\end{figure*}

The IMFs are constructed by counting the number of stars within a mass range of dM = 0.133 in log units. For this, we chose a mass range of 0.2 M$_{\odot}$ to 20 M$_{\odot}$ and divided it into 16 bins. In order to make the selection independent of epoch, we count all the sink particles between one and three free-fall times, with a time resolution of 1/60th $\rm t_{ff}$, and take the average. Alternatively, we have also analyzed the IMFs by selecting the time frames where all the models have converted the same amount of gas mass into sink particles. Either way, we get the same results. The IMFs of the primary models are shown in Fig. \ref{fig:IMFs}. 
In these images, the green dashed lines indicate the Salpeter IMF \citep{1955ApJ...121..161S} and the blue dot-dashed lines represent the Chabrier IMF \citep{2003PASP..115..763C}. The red solid and the purple triple-dotted lines are the best power-law and the best lognormal fits to the data. The IMFs derived from the simulations with (models B and C) or without (model A) magnetic fields differ significantly. The run without magnetic fields resembles the conditions of the Milky Way with an IMF that nicely follows a lognormal shape and has a power-law slope of $\Gamma = -1.38$ above $\rm 1 \,M_{\odot}$. This is also in agreement with earlier numerical studies of the IMF for Milky Way conditions \citep{1997ApJ...486..944E, 2006MNRAS.368.1296B, 2010A&A...522A..24H}. The characteristic mass is slightly larger than expected, i.e., $\rm m_{char} \sim 0.9 \,M_{\odot}$, where the commonly found value in the Milky Way is around $0.3-0.5$ M$_{\odot}$. This could be the result of the strong gravity field in our models. 

For the runs with magnetic fields, the IMF is mainly populated by sub-solar mass stars $\rm \lesssim 0.3 \,M_{\odot}$. While power-law fits below $\rm \sim 1 \,M_{\odot}$ yield slopes of $\rm \Gamma = -2.75$ and $-3.27$, the IMF seems no longer well represented by a single power-law or a lognormal function. Although there is still a log-normal component, its peak is shifted to smaller mass scales, due to the decreased accretion rates discussed in the previous subsection. For model C, the sink particle distribution indeed seems to resemble an exponential form, as the low-mass tail of the IMF is no longer resolved. Besides the lognormal component, magnetic fields tend to create an additional flat component of the IMF above $1-2$ solar masses, as a result of the additional pressure. This second component is a local effect from the magnetic fields, which stabilizes against local collapse until higher densities are reached. The effect seems to occur only locally where the magnetic field strength is particularly high, while most of the formed sinks contribute to the log-normal component in regions dominated by turbulence. This behavior makes sense, considering that the mean mass to flux ratio is still larger than one in all simulations. These findings markedly coincide with the bottom-heavy IMFs observed in massive cluster elliptical galaxies at high redshift \citep{2010Natur.468..940V, 2011ApJ...735L..13V}.

\subsection{An active black hole}
\begin{figure*}[htb!]
\centering
\begin{tabular}{cc}
\begin{minipage}{5.7cm}
\includegraphics[scale=0.36]{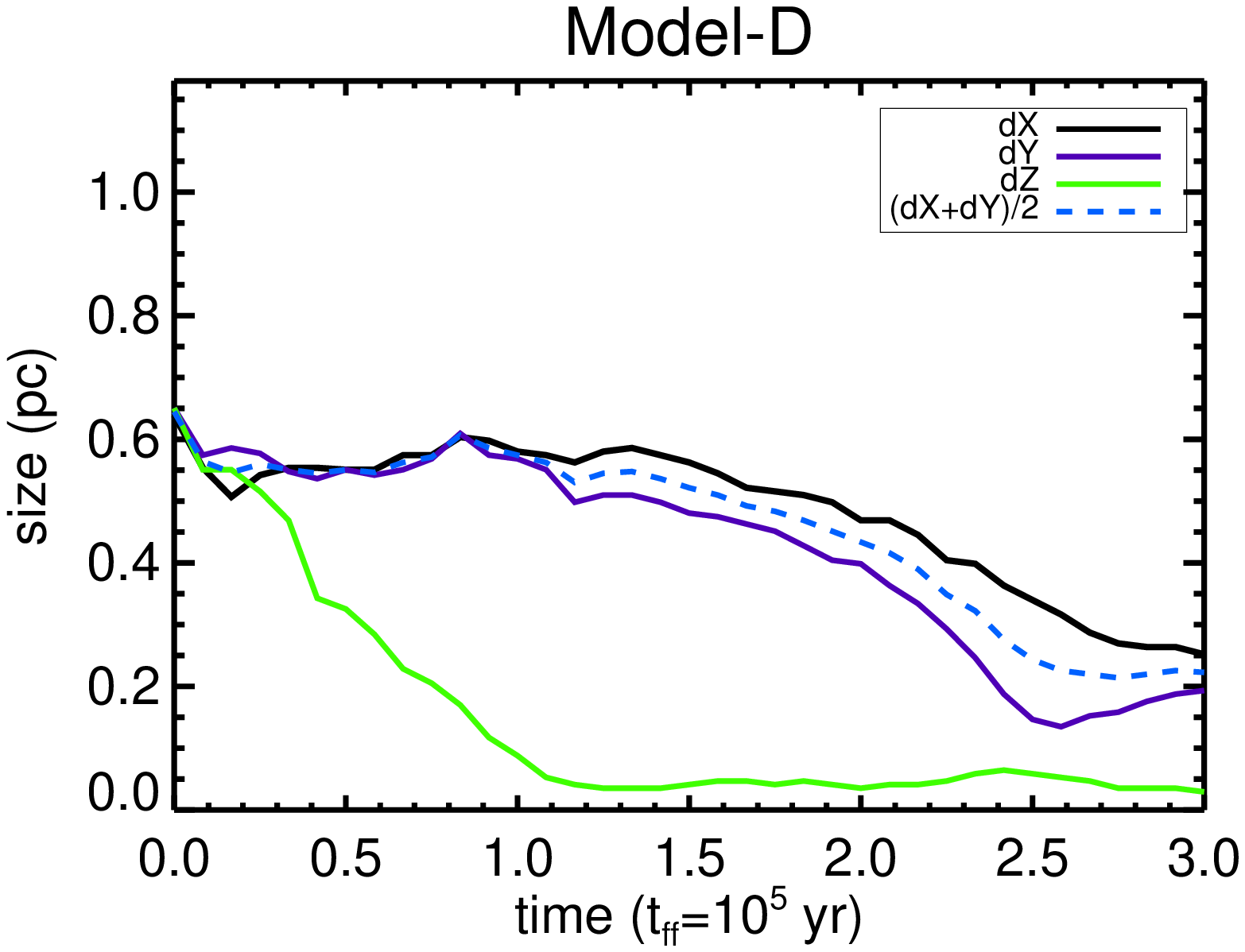}
\end{minipage} &
\begin{minipage}{8cm}
\includegraphics[scale=0.36]{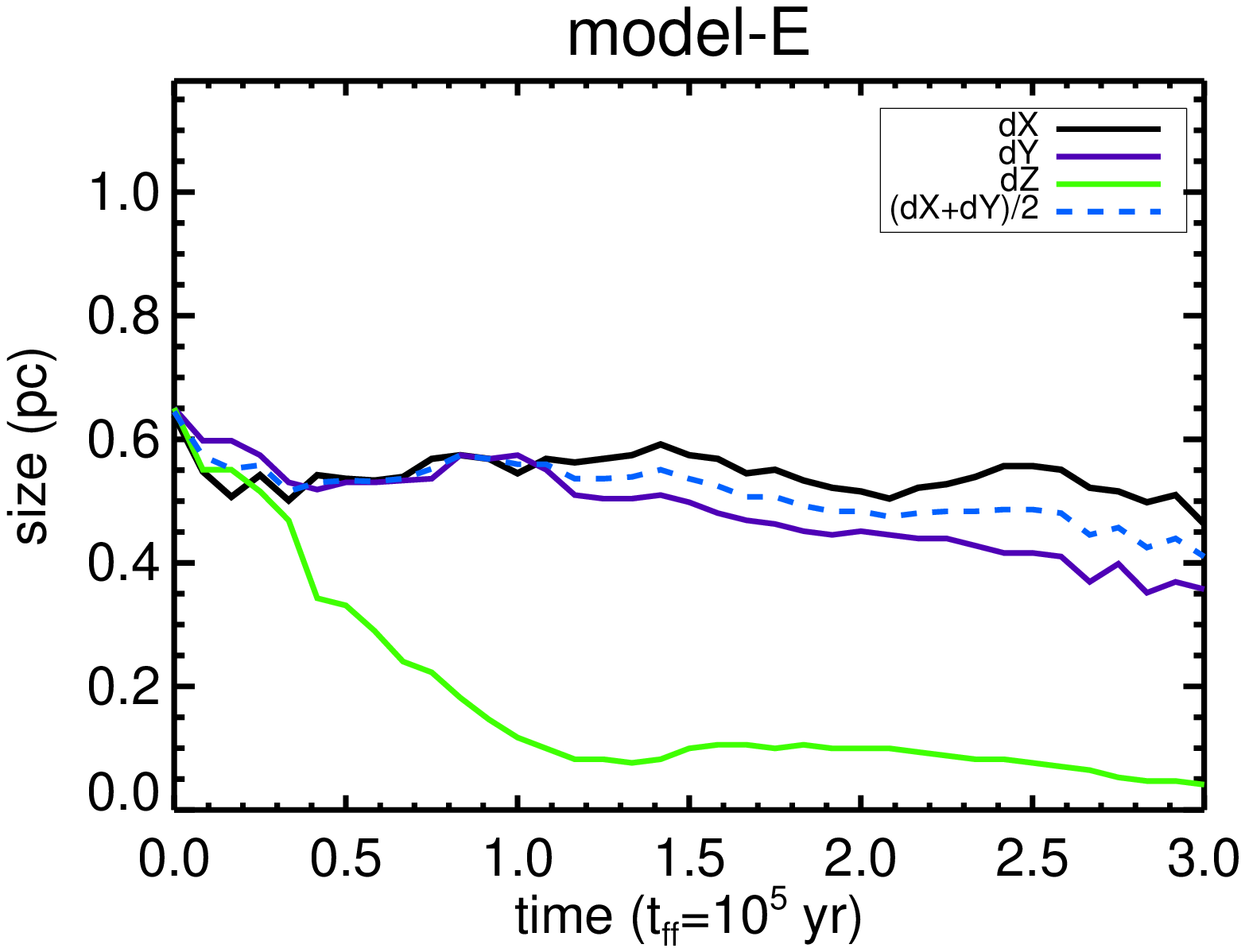}
\end{minipage}
\end{tabular}
\caption{Spatial sizes of the cloud. The size of the cloud in each spatial direction is plotted against time in $\rm t_{ff}$. The sizes in the direction of rotation are denoted by dX and dY, and the height is given by dZ. The dashed line is a representation of the average disk size.}
\label{fig:shapeX}
\end{figure*}

\begin{figure*}[htb!]
\centering
\begin{tabular}{cc}
\begin{minipage}[t]{8.6cm}
\hspace{1.65cm}
\includegraphics[scale=0.36]{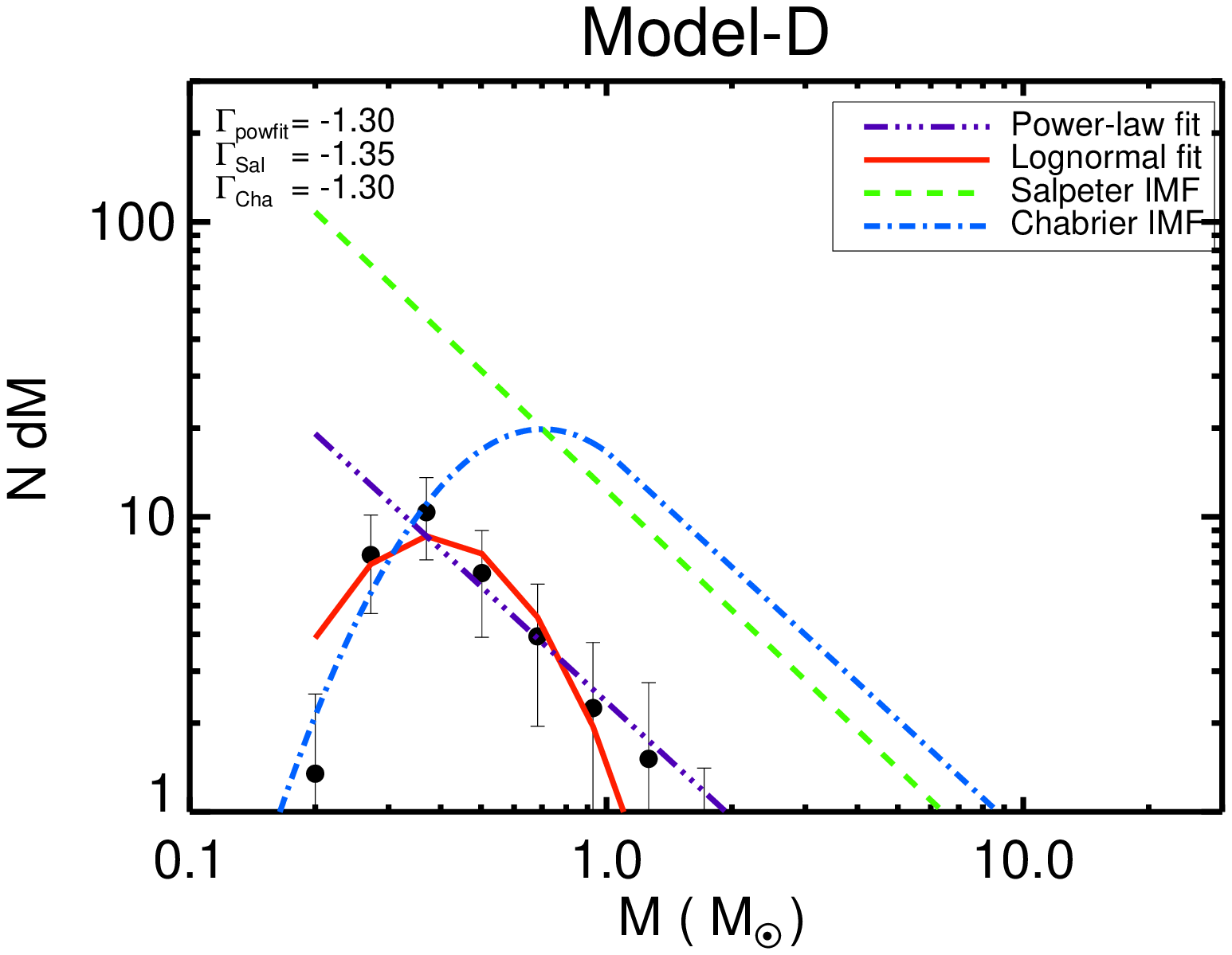}
\caption{The IMF of the X-ray model D. The image displays the time-averaged IMF between 1 and 3 free-fall times, where $\rm t_{ff}=10^{5} \,yr$. In this image, the Salpeter IMF (green dashed) and the Chabrier IMF (blue dot-dashed) are displayed (with arbitrary values along the Y-axis). Two best fits are applied to the data, a power-law fit and a lognormal fit, and are shown as purple and red lines. The slopes above the characteristic mass are given in the upper left corner.}
\label{fig:IMFsX}
\end{minipage}&
\begin{minipage}[t]{8.6cm}
\hspace{-1.25cm}
\includegraphics[scale=0.36]{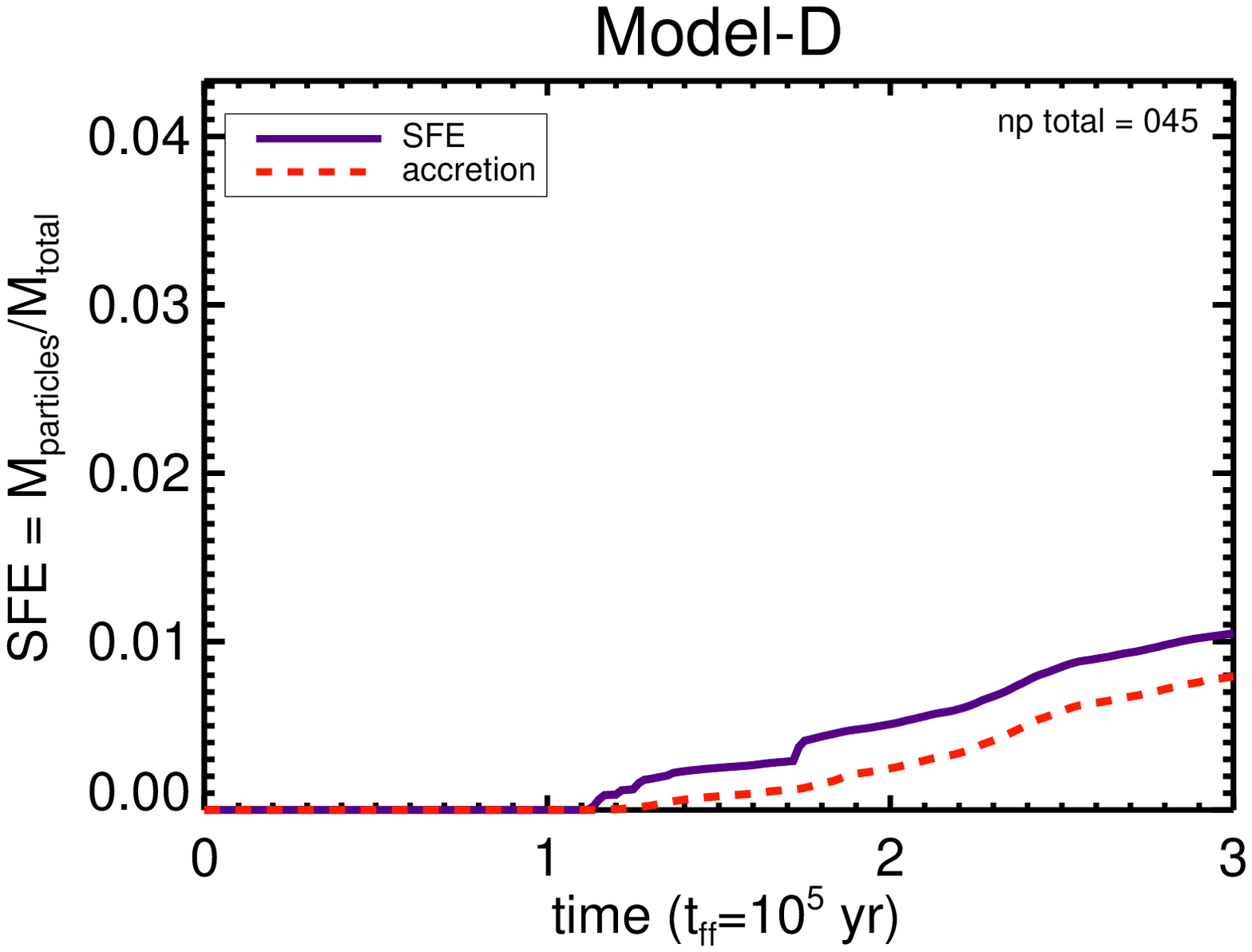}
\caption{The SFE of the X-ray model D. The ratios of the total sink particle mass (solid line) and the total accreted mass (dashed line) with respect to the total mass is plotted against time in $\rm t_{ff}$. The total number of sink particles formed during the run is given in the upper right corner. Note that the plotted range along the Y-axis differs from the range in Fig. \ref{fig:SFEs}.}
\label{fig:SFEsX}
\end{minipage}
\end{tabular}
\end{figure*}

We tested two additional models in which the black hole was considered to be active. The accreting black hole produces strong X-rays that impinge on the model cloud with a flux of 160 erg s$^{-1}$ cm$^{-2}$. The two X-ray models, D and E, have the same parameters as models B and C (Table \ref{tab:parameters}) aside from being irradiated by X-rays. The temperatures in the X-ray models range from 5500 K at the irradiated face of the cloud to 10 K in the most shielded regions. We show the sizes, an IMF, and an SFE in Figs. \ref{fig:shapeX}, \ref{fig:IMFsX}, and \ref{fig:SFEsX}.

If we look at the cloud morphology and compare models D and E (Fig. \ref{fig:shapeX}) against B and C (Fig. \ref{fig:shape}), we see that the sizes are more compact in each direction. This is a direct consequence of X-ray irradiation. X-ray heating increases the external pressure by raising the temperature of the low-density ISM and particularly the cloud surface. This aids collapse. At the same time it also decreases SFEs (Fig. \ref{fig:SFEsX}), because (1) X-ray heating evaporates the outer layers of the cloud, which leads to mass loss, and (2) X-rays increase the temperature inside the cloud in regions where the radiation can penetrate, and by doing so, the Jeans mass is raised. We find that the SFE of model D is only 1\% of the total mass whereas model E shows no star formation at all. For model E, the combined pressure of the magnetic field and the increased internal thermal pressure due to X-rays, where $\rm P_{mag+th} = 2.71 \times 10^{-8} \,dyne \,cm^{-2}$ at $\rm t = 1 \,t_{ff}$, is halting collapse completely, with $\rm P_{grav} = 2.33 \times 10^{-8} \,dyne \,cm^{-2}$ in the cloud core, and a slowly evaporating, quiescent disk forms. In both X-ray models, the cloud loses a significant portion of its mass to the ISM, over 60\% at $\rm t = 1 \,t_{ff}$, especially at densities of $\rm n < 10^{5} \,cm^{-3}$

If we examine the IMF of model D in Fig. \ref{fig:IMFsX}, we see that the IMF has a power-law slope $\Gamma$ of -1.30 and a characteristic mass of 0.37 M$_{\odot}$. The characteristic mass in this model peaks at the same mass as in model B. The global effect of magnetic fields on accretion has the same implications for model D. However, the number of stars is considerably reduced (by a factor 10). We see that X-ray irradiation has strongly quenched star formation in this cloud. The existence of a temperature gradient throughout the cloud is the underlying reason. The formation of low-mass stars is hindered on the irradiated side of the cloud and can only exist in the shielded, cold regions. The more massive stars are affected to a lesser degree. Stars with $\rm M > 2 \,M_{\odot}$ have not formed due to the largely decreased SFE and a flat high-mass contribution is not seen. These effects altogether change the IMF back to a lognormal shape with a Salpeter slope, in a manner conforming with \cite{2002Sci...295...82K}.

If typical clouds have $\rm \sim800\,M_\odot$, as considered here, it implies that predominantly the log-normal component of the IMF will appear, as the overall efficiency is too low for the formation of high-mass stars. In larger clouds, one may however still expect to see the flat component of the IMF as well, provided that the gas reservoir is sufficiently high.

\section{Summary and Conclusions}
\label{sec:conclusion}
We presented a series of numerical simulations to study the impact of magnetic fields on the IMF in star-forming clouds near a black hole. To this end, we created a magnetically supercritical, 800 $\rm M_{\odot}$ cloud at 10 pc from a $\rm 10^{7} M_{\odot}$ black hole and followed its evolution. One model was simulated without magnetic fields as a fixed basis for comparison (model A). This model had similar conditions as the galactic center of the Milky Way. We find that model A results in a universal IMF with typical Salpeter slope of $\Gamma = -1.38$. The characteristic mass of $\rm m_{char} \sim 0.9 \,M_{\odot}$ is somewhat on the large side. We attribute this to the presence of a strong gravitational field from the black hole. To test the impact of magnetic fields, we initiated the cloud with two different magnetic field strengths, 38 and 135 $\mu$G (models B and C). All three models were performed under isothermal conditions.

In all the models, the cloud contracts into a disk within a free-fall time, but the collapse is slower with magnetic fields. We find that the magnetic field lines align parallel to the disk as the disk forms and arrange to an hourglass shape. The alignment of the magnetic field with the disk is because gravity is more important than the magnetic force. The vector components along the XY-plane get enhanced by the increase in density in the course of gravitational collapse, during which the turbulence aids the deformation of the field lines across the molecular cloud. We see less turbulence with increasing magnetic field strength and find that direct turbulent fragmentation is reduced. The first sink particles form along the field lines, but dynamic interactions disperse them quickly. The number of sink particles that form in the magnetic field runs are considerably higher (factor of 2 for $\rm B = 38 \,\mu G$) with smaller masses. The reason is that the slower transport of gas into the center is reducing individual stellar masses and their accretion rates and thus more sink particles can form, while the overall change in cloud morphology, gas reservoir, and turbulence seems to favor enhanced fragmentation. We also notice that there is a jump in the SFEs around 2 $\rm t_{ff}$ in the magnetic field models (Fig. \ref{fig:SFEs}). Sink particles start to accrete more mass at this epoch, because the density increases and mergers are happening in the center of the cloud. 

The IMFs are severely affected by magnetic fields. We find that the IMFs are very bottom heavy with increasing magnetic field strength, but have an additional flat contribution at the high-mass end. Our best fits show very steep power-law slopes below $\sim 1$ M$_{\odot}$, consistent with high redshift observations \citep{2010Natur.468..940V, 2011ApJ...735L..13V}. We also find a decreasing trend in the characteristic mass in the presence of magnetic fields, as a result of the decreasing accretion rates. The respective characteristic masses for models B and C are 0.35 M$_{\odot}$ and 0.20 M$_{\odot}$. A decrease in characteristic mass is also found by \cite{2010ApJ...720L..26L} in their numerical simulations. So overall, both the decreasing stellar masses and the additional flat component in the IMF are in the end due to the magnetic pressure. The first is more because of its global effects, which decreases the accretion rate, and the second is a local effect, which stabilizes against local collapse until a higher mass is reached. \\

\noindent
We summarize our main conclusions for the isothermal magnetic field simulations as follows. When there are magnetic fields:

\begin{itemize}
\item Collapse is slowed down.
\item Star formation occurs along the field lines.
\item Fragmentation is reduced.
\item The number of stars increases by a factor of $1.5-2$.
\item The SFE decreases by $\sim20$\%.
\item The characteristic mass shifts to smaller mass scales.
\item The lognormal IMF is very bottom heavy ($\Gamma \lesssim -2.75$).
\item The IMF has an additional flat high-mass component.
\end{itemize}

It is expected that magnetic fields allow the formation of filaments \citep{2001ApJ...555..178F, 2009MNRAS.398.1082B, 2010ApJ...720L..26L}. We do not see any clear evidence of filaments in our magnetic field simulations. Our different initial conditions might be affecting their formation. Because of the strong gravity of the nearby black hole, the cloud quickly collapses into a disk. The injected turbulence by the tidal shear imposed by the black hole is also influencing the shape of the cloud and likely disallowing filament formation.

The initially chosen magnetic field orientation could be of importance for the results. The orientation of the initial magnetic field vectors in this work is a uniform field with equal strength along each axis. This means that the field lines make an angle of 45 degrees with each axis. We have tested the importance of the initial orientation by running one case with a different orientation, i.e., a uniform magnetic field where the field lines are perpendicular to the orbital plane. There are some differences in results, such as the number of formed sink particles, but we find that the main conclusions (cloud morphology, IMFs, and SFEs) still stand. An in-depth study on the importance of the initial magnetic field orientation is beyond the scope of this work. \\

\noindent
\textit{X-ray feedback by an AGN:} 
We have also simulated a realistic scenario in which the black hole is active and accreting at 10\% Eddington. A large fraction of the AGN's primary output is obscured by interstellar gas and dust close to the accretion disc, but the most energetic wavebands, like X-rays, can penetrate large columns of gas and dust and irradiate the cloud. We have used an X-ray flux of 160 erg s$^{-1}$ cm$^{-2}$, which is typical for obscured AGN environments. This setup is performed with the two aforementioned initial magnetic field strengths of 38 and 135 $\mu$G. \\

\noindent
We summarize our main conclusions for the X-ray models as follows. When there are X-rays:

\begin{itemize}
\item Collapse is aided by the increased external pressure.
\item The cloud slowly evaporates and the Jeans mass is increased.
\item The number of stars is reduced by an order of magnitude.
\item The overall SFE is strongly inhibited.
\item The IMF has a lognormal shape with $\Gamma = -1.30$.
\item For higher-mass clouds, one may expect an additional flat component of the IMF, as seen in the isothermal runs.
\end{itemize}

Our results tell us that in the presence of magnetic fields, star formation in galactic centers experiences a mode where low-mass stars are preferred, but that X-ray feedback reduces the SFE. For $\rm 800\,M_\odot$ clouds as considered here, X-rays make the power-law slope of the IMF flatter and allow a typical Chabrier IMF to emerge. We expect significant changes in the IMF in galactic centers with inactive massive black holes both due to the shifted lognormal as well as the additional flat component, but a decreased SFE and a quenched low-mass star formation in AGN. As observations may have limitations in detecting the low-mass stars, a flat (top-heavy) IMF can be expected in the presence of magnetic fields since, especially at the low-mass end of the IMF, completeness is often an issue, but significant progresses are being made \citep[e.g.,][]{2009ApJ...696..528D}. Evidence for such a component would thus provide strong evidence for the importance of magnetic fields during star formation, as hydrodynamical models generally yield lognormal distributions \citep[e.g.][]{2008ApJ...687..354N}. Increasingly accurate measurements of the IMF in the center of our own galaxy may thus provide a relevant pathway to probe their implications \citep{2007MNRAS.381L..40D, 2009A&A...501..563E, 2012A&A...540A..57H, 2012A&A...540A..14L}.

\begin{acknowledgements}
We are grateful to the anonymous referee for an insightful and constructive report that helped to improve this work. The software used in this work was developed in part by the DOE NNSA ASC- and DOE Office of Science ASCR-supported Flash Center for Computational Science at the University of Chicago. The simulations have been run on the dedicated special purpose machines `Gemini' at the Kapteyn Astronomical Institute, University of Groningen and at the Donald Smits Center for Information Technology (CIT) using the Millipede Cluster, University of Groningen. SH thanks Dongwook Lee for his support on the MHD solver. DRGS thanks for funding through the SPP 1573 `Physics of the Interstellar Medium' (project number SCHL 1964/1-1) and the SFB 963/1 `Astrophysical Flow Instabilities and Turbulence'
\end{acknowledgements}

\bibliography{biblio.fourth.bib}

\begin{thebibliography}{73}
\expandafter\ifx\csname natexlab\endcsname\relax\def\natexlab#1{#1}\fi

\bibitem[{{Aalto}(2005)}]{2005Ap&SS.295..143A}
{Aalto}, S. 2005, \apss, 295, 143

\bibitem[{{Banerjee} {et~al.}(2009){Banerjee}, {V{\'a}zquez-Semadeni},
  {Hennebelle}, \& {Klessen}}]{2009MNRAS.398.1082B}
{Banerjee}, R., {V{\'a}zquez-Semadeni}, E., {Hennebelle}, P., \& {Klessen},
  R.~S. 2009, \mnras, 398, 1082

\bibitem[{{Bec} \& {Khanin}(2007)}]{2007PhR...447....1B}
{Bec}, J. \& {Khanin}, K. 2007, \physrep, 447, 1

\bibitem[{{Beck} {et~al.}(2005){Beck}, {Fletcher}, {Shukurov}, {Snodin},
  {Sokoloff}, {Ehle}, {Moss}, \& {Shoutenkov}}]{2005A&A...444..739B}
{Beck}, R., {Fletcher}, A., {Shukurov}, A., {et~al.} 2005, \aap, 444, 739

\bibitem[{{Bonnell} {et~al.}(2006){Bonnell}, {Clarke}, \&
  {Bate}}]{2006MNRAS.368.1296B}
{Bonnell}, I.~A., {Clarke}, C.~J., \& {Bate}, M.~R. 2006, \mnras, 368, 1296

\bibitem[{{Bourke} {et~al.}(2001){Bourke}, {Myers}, {Robinson}, \&
  {Hyland}}]{2001ApJ...554..916B}
{Bourke}, T.~L., {Myers}, P.~C., {Robinson}, G., \& {Hyland}, A.~R. 2001, \apj,
  554, 916

\bibitem[{{Burgers}(1939)}]{Burgers1939}
{Burgers}, J.~M. 1939, {Mathematical Examples Illustrating Relations Occurring
  in the Theory of Turbulent Fluid Motion}

\bibitem[{{Chabrier}(2003)}]{2003PASP..115..763C}
{Chabrier}, G. 2003, \pasp, 115, 763

\bibitem[{{Clark} {et~al.}(2008){Clark}, {Bonnell}, \&
  {Klessen}}]{2008MNRAS.386....3C}
{Clark}, P.~C., {Bonnell}, I.~A., \& {Klessen}, R.~S. 2008, \mnras, 386, 3

\bibitem[{{Crocker} {et~al.}(2010){Crocker}, {Jones}, {Melia}, {Ott}, \&
  {Protheroe}}]{2010Natur.463...65C}
{Crocker}, R.~M., {Jones}, D.~I., {Melia}, F., {Ott}, J., \& {Protheroe}, R.~J.
  2010, \nat, 463, 65

\bibitem[{{Crutcher}(1999)}]{1999ApJ...520..706C}
{Crutcher}, R.~M. 1999, \apj, 520, 706

\bibitem[{{Crutcher} {et~al.}(1987){Crutcher}, {Troland}, \&
  {Kazes}}]{1987A&A...181..119C}
{Crutcher}, R.~M., {Troland}, T.~H., \& {Kazes}, I. 1987, \aap, 181, 119

\bibitem[{{Da Rio} {et~al.}(2009){Da Rio}, {Gouliermis}, \&
  {Henning}}]{2009ApJ...696..528D}
{Da Rio}, N., {Gouliermis}, D.~A., \& {Henning}, T. 2009, \apj, 696, 528

\bibitem[{{Dib} {et~al.}(2007){Dib}, {Kim}, \&
  {Shadmehri}}]{2007MNRAS.381L..40D}
{Dib}, S., {Kim}, J., \& {Shadmehri}, M. 2007, \mnras, 381, L40

\bibitem[{Dubey {et~al.}(2009)Dubey, Antypas, Ganapathy, Reid, Riley, Sheeler,
  Siegel, \& Weide}]{Dubey2009512}
Dubey, A., Antypas, K., Ganapathy, M.~K., {et~al.} 2009, Parallel Computing,
  35, 512

\bibitem[{{Elmegreen}(1997)}]{1997ApJ...486..944E}
{Elmegreen}, B.~G. 1997, \apj, 486, 944

\bibitem[{{Espinoza} {et~al.}(2009){Espinoza}, {Selman}, \&
  {Melnick}}]{2009A&A...501..563E}
{Espinoza}, P., {Selman}, F.~J., \& {Melnick}, J. 2009, \aap, 501, 563

\bibitem[{{Evans} \& {Hawley}(1988)}]{1988ApJ...332..659E}
{Evans}, C.~R. \& {Hawley}, J.~F. 1988, \apj, 332, 659

\bibitem[{{Falgarone} {et~al.}(2001){Falgarone}, {Pety}, \&
  {Phillips}}]{2001ApJ...555..178F}
{Falgarone}, E., {Pety}, J., \& {Phillips}, T.~G. 2001, \apj, 555, 178

\bibitem[{{Falgarone} {et~al.}(2008){Falgarone}, {Troland}, {Crutcher}, \&
  {Paubert}}]{2008A&A...487..247F}
{Falgarone}, E., {Troland}, T.~H., {Crutcher}, R.~M., \& {Paubert}, G. 2008,
  \aap, 487, 247

\bibitem[{{Federrath} {et~al.}(2010){Federrath}, {Banerjee}, {Clark}, \&
  {Klessen}}]{2010ApJ...713..269F}
{Federrath}, C., {Banerjee}, R., {Clark}, P.~C., \& {Klessen}, R.~S. 2010,
  \apj, 713, 269

\bibitem[{{Federrath} {et~al.}(2011){Federrath}, {Chabrier}, {Schober},
  {Banerjee}, {Klessen}, \& {Schleicher}}]{2011PhRvL.107k4504F}
{Federrath}, C., {Chabrier}, G., {Schober}, J., {et~al.} 2011, Physical Review
  Letters, 107, 114504

\bibitem[{{Fryxell} {et~al.}(2000){Fryxell}, {Olson}, {Ricker}, {Timmes},
  {Zingale}, {Lamb}, {MacNeice}, {Rosner}, {Truran}, \&
  {Tufo}}]{2000ApJS..131..273F}
{Fryxell}, B., {Olson}, K., {Ricker}, P., {et~al.} 2000, \apjs, 131, 273

\bibitem[{{Girart} {et~al.}(2009){Girart}, {Beltr{\'a}n}, {Zhang}, {Rao}, \&
  {Estalella}}]{2009Sci...324.1408G}
{Girart}, J.~M., {Beltr{\'a}n}, M.~T., {Zhang}, Q., {Rao}, R., \& {Estalella},
  R. 2009, Science, 324, 1408

\bibitem[{{Hennebelle} {et~al.}(2011){Hennebelle}, {Commer{\c c}on}, {Joos},
  {Klessen}, {Krumholz}, {Tan}, \& {Teyssier}}]{2011A&A...528A..72H}
{Hennebelle}, P., {Commer{\c c}on}, B., {Joos}, M., {et~al.} 2011, \aap, 528,
  A72

\bibitem[{{Heyer} \& {Brunt}(2004)}]{2004ApJ...615L..45H}
{Heyer}, M.~H. \& {Brunt}, C.~M. 2004, \apjl, 615, L45

\bibitem[{{Hocuk} \& {Spaans}(2010)}]{2010A&A...522A..24H}
{Hocuk}, S. \& {Spaans}, M. 2010, \aap, 522, A24+

\bibitem[{{Hocuk} \& {Spaans}(2011)}]{2011A&A...536A..41H}
{Hocuk}, S. \& {Spaans}, M. 2011, \aap, 536, A41

\bibitem[{{Hosking} \& {Whitworth}(2004)}]{2004MNRAS.347.1001H}
{Hosking}, J.~G. \& {Whitworth}, A.~P. 2004, \mnras, 347, 1001

\bibitem[{{Hu{\ss}mann} {et~al.}(2012){Hu{\ss}mann}, {Stolte}, {Brandner},
  {Gennaro}, \& {Liermann}}]{2012A&A...540A..57H}
{Hu{\ss}mann}, B., {Stolte}, A., {Brandner}, W., {Gennaro}, M., \& {Liermann},
  A. 2012, \aap, 540, A57

\bibitem[{{Kritsuk} {et~al.}(2011){Kritsuk}, {Nordlund}, {Collins}, {Padoan},
  {Norman}, {Abel}, {Banerjee}, {Federrath}, {Flock}, {Lee}, {Li},
  {M{\"u}ller}, {Teyssier}, {Ustyugov}, {Vogel}, \& {Xu}}]{2011ApJ...737...13K}
{Kritsuk}, A.~G., {Nordlund}, {\AA}., {Collins}, D., {et~al.} 2011, \apj, 737,
  13

\bibitem[{{Kroupa}(2002)}]{2002Sci...295...82K}
{Kroupa}, P. 2002, Science, 295, 82

\bibitem[{{Krumholz} {et~al.}(2004){Krumholz}, {McKee}, \&
  {Klein}}]{2004ApJ...611..399K}
{Krumholz}, M.~R., {McKee}, C.~F., \& {Klein}, R.~I. 2004, \apj, 611, 399

\bibitem[{{Larson}(1981)}]{1981MNRAS.194..809L}
{Larson}, R.~B. 1981, \mnras, 194, 809

\bibitem[{Lee \& Deane(2009)}]{Lee:2009:USM:1486277.1486457}
Lee, D. \& Deane, A.~E. 2009, J. Comput. Phys., 228, 952

\bibitem[{Leer(1977)}]{VanLeer1977263}
Leer, B.~V. 1977, Journal of Computational Physics, 23, 263

\bibitem[{{Li} {et~al.}(2010){Li}, {Wang}, {Abel}, \&
  {Nakamura}}]{2010ApJ...720L..26L}
{Li}, Z.-Y., {Wang}, P., {Abel}, T., \& {Nakamura}, F. 2010, \apjl, 720, L26

\bibitem[{{Liermann} {et~al.}(2012){Liermann}, {Hamann}, \&
  {Oskinova}}]{2012A&A...540A..14L}
{Liermann}, A., {Hamann}, W.-R., \& {Oskinova}, L.~M. 2012, \aap, 540, A14

\bibitem[{{Loenen} {et~al.}(2008){Loenen}, {Spaans}, {Baan}, \&
  {Meijerink}}]{2008A&A...488L...5L}
{Loenen}, A.~F., {Spaans}, M., {Baan}, W.~A., \& {Meijerink}, R. 2008, \aap,
  488, L5

\bibitem[{{Maki} \& {Susa}(2004)}]{2004ApJ...609..467M}
{Maki}, H. \& {Susa}, H. 2004, \apj, 609, 467

\bibitem[{{Maki} \& {Susa}(2007)}]{2007PASJ...59..787M}
{Maki}, H. \& {Susa}, H. 2007, \pasj, 59, 787

\bibitem[{{Matsumura} {et~al.}(2012){Matsumura}, {Oka}, \&
  {Tanaka}}]{Matsumura2012}
{Matsumura}, S., {Oka}, T., \& {Tanaka}, K. 2012, ArXiv e-prints

\bibitem[{{Meijerink} \& {Spaans}(2005)}]{2005A&A...436..397M}
{Meijerink}, R. \& {Spaans}, M. 2005, \aap, 436, 397

\bibitem[{{Meijerink} {et~al.}(2007){Meijerink}, {Spaans}, \&
  {Israel}}]{2007A&A...461..793M}
{Meijerink}, R., {Spaans}, M., \& {Israel}, F.~P. 2007, \aap, 461, 793

\bibitem[{{Meijerink} {et~al.}(2011){Meijerink}, {Spaans}, {Loenen}, \& {van
  der Werf}}]{2011A&A...525A.119M}
{Meijerink}, R., {Spaans}, M., {Loenen}, A.~F., \& {van der Werf}, P.~P. 2011,
  \aap, 525, A119+

\bibitem[{{Miyoshi} \& {Kusano}(2005)}]{2005AGUFMSM51B1295M}
{Miyoshi}, T. \& {Kusano}, K. 2005, AGU Fall Meeting Abstracts, B1295+

\bibitem[{{Moss} {et~al.}(2007){Moss}, {Snodin}, {Englmaier}, {Shukurov},
  {Beck}, \& {Sokoloff}}]{2007A&A...465..157M}
{Moss}, D., {Snodin}, A.~P., {Englmaier}, P., {et~al.} 2007, \aap, 465, 157

\bibitem[{{Mouschovias} {et~al.}(2011){Mouschovias}, {Ciolek}, \&
  {Morton}}]{2011MNRAS.415.1751M}
{Mouschovias}, T.~C., {Ciolek}, G.~E., \& {Morton}, S.~A. 2011, \mnras, 415,
  1751

\bibitem[{{Mouschovias} \& {Spitzer}(1976)}]{1976ApJ...210..326M}
{Mouschovias}, T.~C. \& {Spitzer}, Jr., L. 1976, \apj, 210, 326

\bibitem[{{Myers} \& {Gammie}(1999)}]{1999ApJ...522L.141M}
{Myers}, P.~C. \& {Gammie}, C.~F. 1999, \apjl, 522, L141

\bibitem[{{Nakamura} \& {Li}(2005)}]{2005ApJ...631..411N}
{Nakamura}, F. \& {Li}, Z.-Y. 2005, \apj, 631, 411

\bibitem[{{Nakamura} \& {Li}(2008)}]{2008ApJ...687..354N}
{Nakamura}, F. \& {Li}, Z.-Y. 2008, \apj, 687, 354

\bibitem[{{Nakano} \& {Nakamura}(1978)}]{1978PASJ...30..671N}
{Nakano}, T. \& {Nakamura}, T. 1978, \pasj, 30, 671

\bibitem[{{Papadopoulos} {et~al.}(2011){Papadopoulos}, {Thi}, {Miniati}, \&
  {Viti}}]{2011MNRAS.414.1705P}
{Papadopoulos}, P.~P., {Thi}, W.-F., {Miniati}, F., \& {Viti}, S. 2011, \mnras,
  414, 1705

\bibitem[{{P{\'e}rez-Beaupuits} {et~al.}(2007){P{\'e}rez-Beaupuits}, {Aalto},
  \& {Gerebro}}]{2007A&A...476..177P}
{P{\'e}rez-Beaupuits}, J.~P., {Aalto}, S., \& {Gerebro}, H. 2007, \aap, 476,
  177

\bibitem[{{P{\'e}rez-Beaupuits} {et~al.}(2009){P{\'e}rez-Beaupuits}, {Spaans},
  {van der Tak}, {Aalto}, {Garc{\'{\i}}a-Burillo}, {Fuente}, \&
  {Usero}}]{2009A&A...503..459P}
{P{\'e}rez-Beaupuits}, J.~P., {Spaans}, M., {van der Tak}, F.~F.~S., {et~al.}
  2009, \aap, 503, 459

\bibitem[{{Price} \& {Bate}(2008)}]{2008MNRAS.385.1820P}
{Price}, D.~J. \& {Bate}, M.~R. 2008, \mnras, 385, 1820

\bibitem[{{Salpeter}(1955)}]{1955ApJ...121..161S}
{Salpeter}, E.~E. 1955, \apj, 121, 161

\bibitem[{{Schleicher} {et~al.}(2010{\natexlab{a}}){Schleicher}, {Banerjee},
  {Sur}, {Arshakian}, {Klessen}, {Beck}, \& {Spaans}}]{2010A&A...522A.115S}
{Schleicher}, D.~R.~G., {Banerjee}, R., {Sur}, S., {et~al.} 2010{\natexlab{a}},
  \aap, 522, A115+

\bibitem[{{Schleicher} {et~al.}(2010{\natexlab{b}}){Schleicher}, {Spaans}, \&
  {Klessen}}]{2010A&A...513A...7S}
{Schleicher}, D.~R.~G., {Spaans}, M., \& {Klessen}, R.~S. 2010{\natexlab{b}},
  \aap, 513, A7+

\bibitem[{{Seifried} {et~al.}(2011){Seifried}, {Banerjee}, {Klessen}, {Duffin},
  \& {Pudritz}}]{2011MNRAS.417.1054S}
{Seifried}, D., {Banerjee}, R., {Klessen}, R.~S., {Duffin}, D., \& {Pudritz},
  R.~E. 2011, \mnras, 417, 1054

\bibitem[{{Spaans} \& {Meijerink}(2008)}]{2008ApJ...678L...5S}
{Spaans}, M. \& {Meijerink}, R. 2008, \apjl, 678, L5

\bibitem[{{Tomisaka} {et~al.}(1988){Tomisaka}, {Ikeuchi}, \&
  {Nakamura}}]{1988ApJ...335..239T}
{Tomisaka}, K., {Ikeuchi}, S., \& {Nakamura}, T. 1988, \apj, 335, 239

\bibitem[{{van der Werf} {et~al.}(2010){van der Werf}, {Isaak}, {Meijerink},
  {Spaans}, {Rykala}, {Fulton}, {Loenen}, {Walter}, {Wei{\ss}}, {Armus},
  {Fischer}, {Israel}, {Harris}, {Veilleux}, {Henkel}, {Savini}, {Lord},
  {Smith}, {Gonz{\'a}lez-Alfonso}, {Naylor}, {Aalto}, {Charmandaris}, {Dasyra},
  {Evans}, {Gao}, {Greve}, {G{\"u}sten}, {Kramer}, {Mart{\'{\i}}n-Pintado},
  {Mazzarella}, {Papadopoulos}, {Sanders}, {Spinoglio}, {Stacey}, {Vlahakis},
  {Wiedner}, \& {Xilouris}}]{2010A&A...518L..42V}
{van der Werf}, P.~P., {Isaak}, K.~G., {Meijerink}, R., {et~al.} 2010, \aap,
  518, L42+

\bibitem[{{van Dokkum} \& {Conroy}(2010)}]{2010Natur.468..940V}
{van Dokkum}, P.~G. \& {Conroy}, C. 2010, \nat, 468, 940

\bibitem[{{van Dokkum} \& {Conroy}(2011)}]{2011ApJ...735L..13V}
{van Dokkum}, P.~G. \& {Conroy}, C. 2011, \apjl, 735, L13

\bibitem[{{Wada} {et~al.}(2009){Wada}, {Papadopoulos}, \&
  {Spaans}}]{2009ApJ...702...63W}
{Wada}, K., {Papadopoulos}, P.~P., \& {Spaans}, M. 2009, \apj, 702, 63

\bibitem[{{Wang} {et~al.}(2012){Wang}, {Zhang}, {Wu}, {Li}, \&
  {Zhang}}]{2012ApJ...745L..30W}
{Wang}, K., {Zhang}, Q., {Wu}, Y., {Li}, H.-b., \& {Zhang}, H. 2012, \apjl,
  745, L30

\bibitem[{{Wang} {et~al.}(2011){Wang}, {Zhang}, {Wu}, \&
  {Zhang}}]{2011ApJ...735...64W}
{Wang}, K., {Zhang}, Q., {Wu}, Y., \& {Zhang}, H. 2011, \apj, 735, 64

\bibitem[{{Wang} {et~al.}(2010){Wang}, {Li}, {Abel}, \&
  {Nakamura}}]{2010ApJ...709...27W}
{Wang}, P., {Li}, Z., {Abel}, T., \& {Nakamura}, F. 2010, \apj, 709, 27

\bibitem[{{Wolk} {et~al.}(2008){Wolk}, {Bourke}, \&
  {Vigil}}]{2008hsf2.book..124W}
{Wolk}, S.~J., {Bourke}, T.~L., \& {Vigil}, M. 2008, {The Embedded Massive Star
  Forming Region RCW 38}, ed. {Reipurth, B.}, 124--+

\bibitem[{{Yusef-Zadeh} \& {Morris}(1987{\natexlab{a}})}]{1987ApJ...320..545Y}
{Yusef-Zadeh}, F. \& {Morris}, M. 1987{\natexlab{a}}, \apj, 320, 545

\bibitem[{{Yusef-Zadeh} \& {Morris}(1987{\natexlab{b}})}]{1987ApJ...322..721Y}
{Yusef-Zadeh}, F. \& {Morris}, M. 1987{\natexlab{b}}, \apj, 322, 721

\end{thebibliography}

\end{document}